\documentstyle[12pt]{article}

\newcommand{\beq}{\begin{equation}}
\newcommand{\beql}[1]{\begin{equation}\label{eq:#1}}
\newcommand{\eeq}{\end{equation}}
\newcommand{\be}{\begin{equation}}
\newcommand{\ee}{\end{equation}}
\newcommand{\beqn}{\begin{eqnarray}}
\newcommand{\eeqn}{\end{eqnarray}}
\newcommand{\bea}{\begin{eqnarray}}
\newcommand{\eea}{\end{eqnarray}}

\newcommand{\eq}[1]{(\ref{eq:#1})}

\newcommand{\J}{k}
\newcommand{\vol}{{\rm Vol}}
\newcommand{\tf}[2]{dz^{#1}\wedge d\bar{z}^{\bar{#2}}}
\newcommand{\Om}[2]{\Omega^{#1\bar{#2}}}

\newcommand{\Gammab}{\bar{\Gamma}}

\begin{document}
\begin{titlepage}

\title{
	\begin{flushright}
	\begin{small}
	PUPT-1690\\
	UPR-744-T \\
	ILL-(TH)-97-3\\
	hep-th/9704143\\
	\end{small}
	\end{flushright}
	\vspace{1.cm}
	Branes at Angles and Black Holes 
	}

\author{
Vijay Balasubramanian
\thanks{vijayb@pauli.harvard.edu}\\
	\small Lyman Laboratory of Physics\\
	\small Harvard University\\
	\small Cambridge, MA 02138 \\
\and
Finn Larsen
\thanks{larsen@cvetic.hep.upenn.edu}\\
	\small David Rittenhouse Laboratories\\
	\small University of Pennsylvania\\
	\small Philadelphia, PA 19104 \\
\and\\
Robert G. Leigh
\thanks{rgleigh@uiuc.edu} \\
	\small Department of Physics\\
	\small University of Illinois at Urbana-Champaign\\
	\small Urbana, IL 61801.
}

\date{ }
\maketitle

\begin{abstract}
We construct the most general supersymmetric configuration of
D2-branes and D6-branes on a 6-torus. It contains arbitrary numbers of
branes at relative $U(3)$ angles. The corresponding supergravity
solutions are constructed and expressed in a remarkably simple form,
using the complex geometry of the compact space. The spacetime
supersymmetry of the configuration is verified explicitly, by solution
of the Killing spinor equations, and the equations of motion are
verified too. Our configurations can be interpreted
as a 16-parameter family of regular extremal black holes in four
dimensions. Their entropy is interpreted microscopically by counting
the degeneracy of bound states of D-branes.  Our result agrees in
detail with the prediction for the degeneracy of BPS states in terms
of the quartic invariant of the E(7,7) duality group.
\end{abstract}

\end{titlepage}
\newpage

\section{Introduction}

In recent work the description of the $p$-brane solitons of supergravity
theory at weak coupling as D-branes in string theory has been exploited
to elucidate many nonperturbative aspects of string theory. An important
strategy has been the construction of complicated solutions to the
supergravity equations of motion by combination of simpler building
blocks that can be unambiguously identified with individual 
branes. (See \cite{gppkt} for example.) Such
composite structures are naturally identified with bound states of the
elementary constituents. A striking success of this reasoning has been
the representation of a large class of extremal black holes as bound
states of p-branes. This has led to the interpretation of the entropy of
these black holes in terms of the bound state degeneracy of the
corresponding system of
branes~\cite{sv1,structure,callan96a,strom96b,dvv1}.

However, the early study of bound states of $p$-branes was limited
because it was only possible to construct supersymmetric configurations
corresponding to branes that were {\em orthogonal} to each other.  It is
now known that in fact there are also supersymmetric bound states where
the component branes are at relative {\em angles}~\cite{bdl,bl}. Very
recently, explicit classical solutions to the supergravity equations
that correspond to some of these configurations of branes at angles have
been presented in the
literature~\cite{gauntlett,behrndt,mcgill1,costa,hambli}.

In this paper, we construct configurations of branes which cover
extended regions of the parameter space of branes at angles. We begin,
in Sec.~\ref{sec:susy}, by considering Type IIA string theory
compactified on the 6-torus  $T^6$, and construct a new class of
supersymmetric bound states of D2-branes oriented at arbitrary $U(3)$
angles in the presence of an arbitrary number of D6-branes. Previous
constructions involving D2-brane at angles are a subset of those
presented here~\cite{bdl,bl}. Indeed, we find the most general class of
supersymmetric bound states of D2-branes and D6-branes permissible on
$T^6$.

In Sec.~\ref{sec:ansatz} we present the corresponding classical
solutions. They have a remarkably simple structure that becomes
apparent when they are displayed in terms of the complex geometry of
the compact space: branes at angles are simply described by the
superposition of two-forms aligned with the constituent branes, with
the superposition coefficients taken to be harmonic functions on the
noncompact space.  The two-form resulting from the superposition
characterizes the ensemble of branes and enters the internal metric as
a simple modification of its K\"{a}hler form.  This result generalizes
the ``harmonic function rule'' for construction of orthogonally
intersecting branes to a setting involving branes at
angles~\cite{harm}. It is remarkable that the general case presented
here may be cast in such a simple form, simpler than the special cases
known hitherto. In Sec.~\ref{sec:eom} it is verified explicitly that
our configurations indeed solve the Killing spinor equations of
supergravity. This requires a strategy that takes advantage of
both the diagonal spacetime metric and the complex geometry of
the compact space. The explicit verification of the equations of
motion, presented in Appendix~\ref{app:eom}, similarly uses the
complex geometry heavily. It is apparent that our construction 
applies to
black holes on general Calabi-Yau manifolds, and we will 
develop the details in a forthcoming article~\cite{us}.

Generically our classical solutions can be interpreted as regular
four-dimensional black holes with finite horizon area. Indeed this is
the main motivation for our investigation. The most general black hole
that couples to electric RR 3-form and RR 7-form gauge fields is
parametrized by 16 charges: the projections of the 2-form charges on
the 15 2-cycles of the 6-torus, and the 6-form charge. The black holes
that we construct account for the complete 16 dimensional phase space.
T-dualizing the entire $T^6$ converts the D2-branes and D6-branes into
D4-branes and D0-branes allowing us to analyze another complete 16
parameter subspace of the extremal black hole phase space. The most
general Type IIA black hole carrying purely Ramond-Ramond charges can
be understood as an amalgam of these two constructions. Their
combination involves fluxes on the D-brane
world-volumes. We leave the general analysis of classical solutions
corresponding to D-branes with fluxes to a future paper~\cite{us} and
conclude this work in Sec.~\ref{sec:entropy} by showing that the
entropy of our general black holes can be accounted for
microscopically in terms of the corresponding D-brane bound state
degeneracy.

\section{D-branes at angles}
\label{sec:susy}

In this section we use the techniques of~\cite{bdl,bl} to
construct supersymmetric configurations of D2-branes
and D6-branes on a 6-torus. We find that a general 
supersymmetric state of D2-branes on $T^6$ can be constructed by placing 
an arbitrary number of branes at relative $U(3)$ 
angles\footnote{The results in this section are also valid
before compactification. In general the branes of such 
configurations are localized in the internal dimensions. 
We refer to the compactified case to ease the comparison with 
results presented in later sections.}. 
Then we show that a D6-brane can always be added to such a 
configuration without breaking supersymmetry. 
T-dualizing the entire torus yields the general supersymmetric 
state of D4-branes and D0-branes on $T^6$. 
Finally, an explicit example with three D2-branes 
and a D6-brane is developed in detail, and special cases are
compared with results from the literature.  

\subsection{Preliminaries}
In this section we work in light-cone frame where the two
supersymmetries of Type II string theory, $Q$ and $\tilde{Q}$, are
16-component chiral $SO(8)$ spinors.  A D$p$-brane imposes the
following projection on $Q$ and $\tilde{Q}$~\cite{dnotes,bdl,bl}
\footnote{We take $\Gamma_{11} \tilde{Q} = + \tilde{Q}$.}:
\begin{equation}
Q \pm \Omega_p(\gamma) \tilde{Q}= 0
\label{eq:susyproj}
\end{equation}
Here $\Omega$ is the volume form of the brane:
\begin{equation}
\Omega_p(\gamma) = {1\over (p+1)!}\epsilon_{i_0\cdots i_p} 
\gamma^{i_0} \cdots \gamma^{i_p} 
\label{eq:Omega}
\end{equation}
Note that this operator is normalized to be a projection operator.
The $\pm$ signs in Eq.~\ref{eq:susyproj} distinguish between branes and
anti-branes or, equivalently, between opposite orientations of the
D$p$-brane. For simplicity, we choose the moduli of the torus to be
unity, {\it i.e.} the torus is square\footnote{Our treatment may be
extended to a general torus by using a vielbein as in~\cite{bl}.}.

We are interested in configurations composed of many D-branes.  Here
supersymmetry demands that the projections imposed by each brane are
simultaneously satisfied.  To solve the resulting system of equations it is
convenient to introduce complex coordinates $(z_1,z_2,z_3)$ on the
6-torus which are  related to the real coordinates as
$z_\mu=(y_{2\mu-1}+iy_{2\mu})/\sqrt{2}$, $\mu=1,2,3$. The
corresponding complexified Gamma matrices are
$\Gamma^{\mu}=(\gamma^{2\mu-1}+i\gamma^{2\mu})/2$ and their complex
conjugates $\Gammab^{\bar \mu}=(\gamma^{2\mu-1}-i\gamma^{2\mu})/2$.
These matrices obey a Clifford algebra\footnote{We use a mostly
positive metric everywhere in this paper.}:
\begin{equation}
\{\Gamma^{\mu}, \Gamma^{\nu} \} = 
\{\Gammab^{\bar\mu},\Gammab^{\bar\nu} \} =0~~~;~~~~~
\{ \Gamma^\mu, \Gammab^{\bar\nu} \} =
\delta^{\mu{\bar\nu}}
\label{eq:cliff1}
\end{equation}
Taking $z_4$ and $\bar{z}_4$ to be complex coordinates describing
the remaining two transverse directions in the light-cone frame we
also define the corresponding complexified Gamma matrices $\Gamma^4$
and $\bar{\Gamma}^{\bar{4}}$.  These $\Gamma$ matrices also obey a
Clifford algebra with the $\Gamma^\mu$ in Eq.~\ref{eq:cliff1}.  

The 16-component SO(8) chiral spinors $\tilde{Q}$ can  be
represented in terms of a Fock basis $|n_1,n_2,n_3\rangle \otimes |
n_4\rangle$ on which the $\Gamma_\mu$ ($\Gammab_{\bar\mu}$) and
$\Gamma_4$ ($\bar{\Gamma}_{\bar{4}}$) act as annihilation (creation)
operators.  Specifically:
\begin{equation}
\Gammab_{\bar\mu}\Gamma_{\mu}|n_1,n_2,n_3\rangle \otimes |n_4\rangle
=n_\mu|n_1,n_2,n_3\rangle \otimes |n_4 \rangle 
\end{equation}
where the $n_\mu$ and $n_4$ take values $0$ and $1$. In this basis it
will be simple to find spinors that satisfy the projections imposed by
the branes.  It will be useful to know how $\tilde{Q}$ transforms
under the $SO(6) \subset SO(8)$ rotations acting on the 6-torus on
which the branes are wrapped.   Since $|n_4\rangle$
is inert under these rotations, the 16-component $SO(8)$ spinor
$\tilde{Q}$ transforms in the 8-dimensional spinor representation of
$SO(6)$.

\subsection{2-branes at $U(3)$ angles}

We will consider an arbitrary number of 2-branes wrapped around
various cycles.  Choose a coordinate system so one of the cycles is
the $(y_1, y_2)$ cycle. According to Eq.~\ref{eq:susyproj} the 2-brane on
this cycle imposes the projection $Q +\gamma^0\gamma^1\gamma^2
\tilde{Q} = 0$. We write this as:
\begin{equation}
\gamma^0 Q =  \gamma^1 \gamma^2 \tilde{Q} = 
-i(2\Gammab^{\bar 1}\Gamma^1 -1)\tilde{Q} 
\label{eq:susyrefconfig1}
\end{equation}
Let us consider a collection of 2-branes rotated relative to this
reference brane on the 6-torus.  The ith brane is rotated by some
$R_i \in SO(6)$ and imposes the supersymmetry
projection:
\begin{equation}
\gamma^0 Q =  (R_i\gamma)^1 (R_i\gamma)^2 \tilde{Q}
\end{equation}
where $R_i$ is in the fundamental representation of $SO(6)$.  (Of
course, on the 6-torus branes can only be rotated by a discrete
subroup of $SO(6)$ if we require that they have finite winding number
on all cycles.) Let ${\cal S}_{(R_i)}$ denote the corresponding
rotation matrix in the spinor representation of $SO(6)$. Then the ith
brane imposes the projection:
\begin{equation}
\gamma^0 Q = -i {\cal S}_{(R_i)} \: (2\Gamma^{\bar 1}\Gamma^1-1) \: 
{\cal S}^{\dagger}_{(R_i)} \:
\tilde{Q}
\label{eq:susyrotat1}
\end{equation}
The Fock space elements $|n_1 n_2 n_3 \rangle \otimes |n_4\rangle$
which form a basis for the spinors $\tilde{Q}$ are eigenstates of the
$\Gamma$-matrix projections in Eq.~\ref{eq:susyrefconfig1} and
Eq.~\ref{eq:susyrotat1}: $-i(2\Gammab^{\bar j} \Gamma^j - 1) |n_1 n_2
n_3 \rangle \otimes |n_4 \rangle = i(1 - 2n_j) |n_1 n_2 n_3 \rangle
\otimes |n_4\rangle$.  Therefore, there are simultaneous solutions of
Eq.~\ref{eq:susyrefconfig1} and all the Eqs.~\ref{eq:susyrotat1} for
each $i$, if there exist some $\tilde{Q}$ which are singlets under
{\em all} the rotations: ${\cal S}_{(R_i)}\tilde{Q} = {\cal
S}^{\dagger}_{(R_i)}\tilde{Q} = \tilde{Q}$.

     Given such a collection of $R_i \in SO(6)$ that leave some
$\tilde{Q}$ invariant, it is clear that any product of the $R_i$ and
their inverses will have the same property.  It is easily seen that the set
of such products will be a subgroup of $SO(6)$.  The problem of
finding supersymmetric relative rotations is therefore reduced to one
of finding subgroups of $SO(6)$ that leave some $\tilde{Q}$
invariant. As discussed in the previous subsection, $\tilde{Q}$
transforms in the 8-dimensional spinor representation of $SO(6)$.  The
largest subgroup of $SO(6)$ under which spinors transform as singlets
is $SU(3)$ with the decomposition ${\bf 8} \rightarrow {\bf 1 + 3 +
\bar{3} + 1}$. This tells us that in general an arbitrary collection
of branes that are related by $SU(3)$ rotations is supersymmetric.  In
fact, the branes can be related by $U(3)$ rotations and still be
supersymmetric because the $U(1)$ factor in $U(3) = SU(3) \times U(1)$
cancels between ${\cal S}_{(R_i)}$ and ${\cal S}^\dagger_{(R_i)}$ in
Eq.~\ref{eq:susyrotat1}.  In the special case when only two branes are
present, the relative rotation can always be represented as an element
of $SO(4)$ in the four dimensions spanned by the branes.  In that case
supersymmetry is preserved when the rotation is in an $SU(2)$ subgroup
of $SO(4)$ as discussed in~\cite{bdl,bl}.  However, when three or 
more branes are present they can explore all six dimensions of the
torus and the present analysis applies.

    We can determine the amount of supersymmetry surviving the
presence of $U(3)$ rotated D2-branes by looking for $U(3)$ invariant
spinors $\tilde{Q}$.  Given the reference configuration
Eq.~\ref{eq:susyrefconfig1} and the Fock basis discussed above, it is
readily shown that the $U(3)$-invariant spinors are $\tilde{Q} = \{
|000\rangle \otimes |n_4\rangle, |111 \rangle \otimes |n_4\rangle\}$
where $n_4=\{0,1\}$.  These four solutions give the equivalent of
$N=1, d=4$ supersymmetry. (We will give a detailed explicit example
that illustrates this general discussion in Sec.~\ref{sec:2226susy}.)

Note that we can have an {\em arbitrary} number of D2-branes rotated at
arbitrary $U(3)$ angles on the torus. In the special case that the
rotations are in an $SU(2)$ subgroup of $U(3)$, we have enhanced
supersymmetry since there are more spinors that are invariant under
these rotations.  Under $SU(2)$, the 8-dimensional spinor of $SO(6)$
decomposes as ${\bf 1 + 2 + 2+ 1 + 1 + 1}$.  Recalling that the
16-component $SO(8)$ spinor $\tilde{Q}$ has a Fock space expansion
$|n_1,n_2,n_3\rangle \otimes |n_4\rangle$ where only the
$|n_2,n_2,n_3\rangle$ tranforms under $SO(6)$ we conclude that there
are eight $SU(2)$ invariant solutions for $\tilde{Q}$.  This gives the
equivalent of $N=2, d=4$ supersymmetry, and reduces to the analysis of
$SU(2)$ rotated 2-branes in~\cite{bdl,bl}.

The significance of the $U(3)$ subgroup of $SO(6)$ is that it
preserves some complex structure of the torus.
Explicitly, there are coordinates $(z_1, z_2, z_3)$ of the 6-torus
that transform in the ${\bf 3}$ of $U(3)$ as:
\begin{equation}
\pmatrix{z_1\cr z_2\cr z_3}\to R\pmatrix{z_1\cr z_2\cr z_3}
\label{uthreerot}
\end{equation}
where $R$ is in the fundamental representation of $U(3)$.  The $z_i$
are associated with complexified matrices $\Gamma^i$ as discussed
above, and these transform in the spinor representation as
$(R\Gamma)^i \rightarrow {\cal S}\Gamma^i {\cal S}^{\dagger}$.  In the
preceding analysis we have essentially found that a system of
2-branes that are wrapped on arbitrary $(1,1)$ cycles relative to
a given complex structure will be supersymmetric.  Essentially, the
choice of a pair of supersymmetric branes at angles picks out a
distinguished complex structure for the torus and any number of
further branes can be wrapped on $(1,1)$ cycles relative to that
complex structure.  Since any two $(1,1)$ cycles are related by $U(3)$
rotations, this means that the branes are $U(3)$ rotated {\em
relative} to each other.

  We can also consider the effects of {\em global} rotations on the
entire system of branes.  It is clear that global rotations of all the
branes together will not affect the analysis of supersymmetry.  Global
rotations that belong to $U(3) \subset SO(6)$ are already included in
the class of configurations of branes at {\em relative} $U(3)$ angles.
So only the remaining global $SO(6)/U(3)$ acts nontrivially to produce
new configurations.  Indeed, by definition, $SO(6)/U(3)$ rotates the
nine $(1,1)$ cycles of the torus that are preserved by the $U(3)$ into
the six $(2,0)$ and $(0,2)$ cycles.  As discussed above, an arbitrary
number of branes at relative $U(3)$ angles can be wrapped
supersymmetrically along any $(1,1)$ cycles.  Manifestly, the global
$SO(6)/U(3)$ rotates these branes into the remaining $(0,2)$
and $(2,0)$ cycles of the torus.
In Sec.~\ref{sec:udual} we will argue that the most general BPS
configuration of 2-branes on a 6-torus can be constructed by applying
these global $SO(6)/U(3)$ rotations to our 2-branes at $U(3)$ angles.

\subsection{Adding D6-branes}
An arbitrary number of D6-branes can be added to the system of
$U(3)$-rotated D2-branes discussed above without breaking
any additional supersymmetry. 
Recall that the $\pm$ sign in the projection condition
Eq.~\ref{eq:susyproj} reflects the two possible orientations of a
brane.  The presence of the $U(3)$ rotated D2-branes
selects the D6-brane orientation: only D6-branes with the orientation
associated with the minus sign in Eq.~\ref{eq:susyproj} can be introduced
without breaking supersymmetry. With this orientation the
supersymmetry projection becomes:
\begin{equation}
\gamma^0 Q = -i(2\Gammab^{\bar 1}\Gamma^1 - 1)
(2\Gammab^{\bar 2}\Gamma^2 - 1)
(2\Gammab^{\bar 3}\Gamma^3 - 1) 
\tilde {Q}
\label{eq:6proj}
\end{equation}
The spinors $\tilde{Q} = \{ |000\rangle, |111\rangle \}$ provide
simultaneous solutions to all the $U(3)$-rotated D2-brane 
conditions Eq.~\ref{eq:susyrotat1}. It is immediately recognized that 
they also solve the D6-brane condition Eq.~\eq{6proj}, giving the
same relation between $Q$ and ${\tilde Q}$.

We will see later that the corresponding classical
configurations are four dimensional black
holes  with finite area. It is clear from the construction
here that such black holes can be interpreted as bound states of
D6-branes with the $U(3)$-rotated 2-branes.

\subsection{D0-branes and D4-branes}

Given the configuration constructed above, we may produce a similar
configuration involving only D0-branes and D4-branes by T-dualizing
the entire $T^6$. This transformation inverts the volume of the
manifold and exchanges D6-branes with D0-branes.  T-duality also
converts each of the D2-branes in the system into a D4-brane wrapped
on the $(2,2)$-cycle orthogonal to the $(1,1)$-cycle on which the
D2-brane is wrapped.  It follows that an arbitrary number of D4-branes
wrapped on $T^6$ will be supersymmetric if they are related by
relative $U(3)$ rotations of the complex coordinates.  For the reasons
discussed above, these $U(3)$-rotated systems are the most general
supersymmetric configurations of D4-branes on $T^6$.

\subsection{Example: 2226 with branes at angles}
\label{sec:2226susy}

In this subsection we will work out an explicit example of D2-branes
at $U(3)$ angles in the presence of a D6-brane.  Three D2-branes are
arranged as follows: one  along $y_1$ and $y_2$, another
along $y_1 \cos\alpha - y_3 \sin\alpha$ and $y_2 \cos\alpha -
y_4\sin\alpha$, and the third one along $y_1 \cos\beta - y_5
\sin\beta$ and $y_2 \cos\beta - y_6\sin\beta$.  In this system the
second and third branes are rotated relative to the first one by
independent $SU(2) \subset U(3)$ rotations.  However, $U(3)$ does not support
several commuting $SU(2)$ subgroups; so clearly this configuration
explores  $U(3)$ in a nontrivial way although the individual branes
are separately rotated only by $SU(2)$

The supersymmetry projections implied by the three
D2-branes and the D6-brane are:
\begin{eqnarray}
\gamma^0 Q &=& \gamma^1 \gamma^2 \tilde{Q} \nonumber \\
\gamma^0 Q &=& \left(\gamma^1 \cos\alpha - \gamma^3 \sin\alpha \right)
      \left(\gamma^2 \cos\alpha - \gamma^4 \sin\alpha \right) 
      \tilde{Q} \nonumber \\
\gamma^0 Q &=&  \left(\gamma^1 \cos\beta - \gamma^5 \sin\beta \right)
      \left(\gamma^2 \cos\beta - \gamma^6 \sin\beta \right) 
      \tilde{Q} \nonumber \\
\gamma^0 Q &=& -\gamma^1 \cdots \gamma^6 \tilde{Q} \nonumber 
\end{eqnarray}
Defining as before
the complexified $\Gamma$-matrices,the supersymmetry conditions can be 
written:
\begin{eqnarray}
\gamma^0 Q &=& -i(2\Gammab^{\bar 1}\Gamma^1-1)\tilde{Q} \nonumber \\
\gamma^0 Q &=& -i[(2\Gammab^{\bar 1}\Gamma^1-1)\cos^2\alpha + 
(2\Gammab^{\bar 2}\Gamma^2-1)\sin^2\alpha
        + 2\sin\alpha\, \cos\alpha (\Gamma^1 \Gammab^{\bar 2}
                                     + \Gamma^2 \Gammab^{\bar 1} )
   ] \tilde{Q}   \nonumber \\
\gamma^0 Q &=& -i[(2\Gammab^{\bar 1}\Gamma^1-1)\cos^2\beta + 
(2\Gammab^{\bar 3}\Gamma^3-1)\sin^2\beta
        + 2\sin\beta\, \cos\beta (\Gamma^1 \Gammab^{\bar 3}
                                     + \Gamma^3 \Gammab^{\bar 1} )
   ]\tilde{Q} \nonumber \\
\gamma^0 Q &=& -i(2\Gammab^{\bar 1}\Gamma^1-1)
(2\Gammab^{\bar 2}\Gamma^2-1)
(2\Gammab^{\bar 3}\Gamma^3-1) \tilde{Q} \nonumber 
\end{eqnarray}
This system of equations has two solutions: $\Gamma^\mu {\tilde Q}=0$,
for which all of them reduce to $\gamma^0 Q=i{\tilde Q}$ ; and
$\Gammab^{\bar\mu} {\tilde Q}=0$, for which they become $\gamma^0
Q=-i{\tilde Q}$. Each of the two solutions imposes 4 projections; so
the complete configuration preserves $2\times 1/16=1/8$ of the
supersymmetry of the vacuum.  As expected the preserved spinors can be
written in the Fock basis as $\tilde{Q} = \{|000\rangle, |111\rangle
\}$.

If we eliminate the D6-brane and the third D2-brane, 
we are left with two D2-branes at an SU(2) angle. This is the simplest
non-trivial example of branes at angles. It was analyzed in detail in
Refs.~\cite{bdl,bl}. Another special case is $\alpha=\beta={1\over 2}\pi$ 
where we find the more familiar configuration of three orthogonally 
intersecting D2-branes in a D6-brane background. This was analyzed
in Ref.~\cite{vf}.

\section{Classical Branes at Angles}
\label{sec:ansatz}

In Sec.~\ref{sec:susy}, we saw that the most general supersymmetric
configuration of D2-branes on $T^6$ consists of a collection of branes
at $U(3)$ rotations. In other words all the D2-branes are wrapped on
$(1,1)$-cycles relative to some given complex structure.  In this
section we will write the corresponding classical solutions to the
supergravity equations.  Various special cases of our solutions and
their M-theory interpretations have been discussed in~\cite{gppkt} and
subsequent publications.  We consider the case where the asymptotic
6-torus is square and has unit moduli.  The solutions are most easily
written in terms of the complex geometry of the compact space and so
we begin by defining notation.

Choosing complex coordinates $z^j = (x^{2j-1} + ix^{2j})/\sqrt{2}$, 
the K\"ahler form of the asymptotic torus is:
\begin{equation}
k=i\sum_{J=1}^3 \tf{J}{J}
\end{equation}
The volume of the asymptotic torus is:
\begin{equation}
\vol(T^6)=\int_{T^6}{\rm dVol} = \int_{T^6} {\J\wedge\J\wedge\J\over 3!}
\end{equation}
and can been set equal to $1$  without loss of generality by taking
the asymptotic moduli to be unity. Our conventions for the normalization
of forms, wedge products and Hodge dual may be found in Appendix A.

Now consider a collection of 2-branes wrapped on a square 6-torus and
let $\omega_j$ be the volume form corresponding to the $(1,1)$-cycle on
which the $j$th brane is wrapped.  In keeping with the notation of
Sec.~\ref{sec:susy} we think of each brane as being $U(3)$ rotated with
respect to a reference configuration on the $(z^1,\bar{z}^1)$ torus.
This gives:
\begin{equation}
\omega_j=i \, (R_{(j)})^1_J \, (R^*_{(j)})^1_K \, \tf{J}{K}.
\end{equation}
Branes wrapped at angles on the torus in this way will act as sources
for the geometry causing the moduli to flow between infinity and 
the position of
the branes in the classical solution.  Remarkably, it turns out that to
understand how the geometry behaves it is sufficient to simply add the
$(1,1)$-forms of each 2-brane, with coefficients that are 
harmonic functions on the non-compact transverse 4-space:
\begin{equation}
\omega = \sum_j X_j \,\omega_j
\end{equation}
In general we can choose $X_j = P_j/|\vec{r} - \vec{r}_j|$ where
$\vec{r}_j$ is the position of the jth brane and $\vec{r}$ is the
coordinate vector in the noncompact space.  When the $\vec{r}_j$ are
all different from each other, the constituent branes of the solution
are separated in the noncompact space and can be easily distinguished.
In our case, we will be principally interested in four-dimensional
black hole solutions and so we will usually work explicitly with the
one-center form:
\begin{equation}
X_j = {P_j \over r}
\end{equation}
where $P_j$ is the charge carried by the $j$th 2-brane\footnote{To the
extent that we are interested only in cohomology (it is natural to
consider a basis of minimal area cycles), $\omega$ is an element of
$H^{(1,1)}(T^6,{\cal O})$, with ${\cal O}$ the appropriate space of
harmonic functions.  However, because of the quantization condition on
the $P_j$'s (see Sec.~\ref{sec:quant}), this is essentially integer
cohomology.}.  Note again that we are only choosing the one-center
form for the discussion of four dimensional black holes.  The proof of
spacetime supersymmetry in Sec. 6.4, for example, goes through for the
general multi-center form of $X_j$.

In the one-center case, after quantizing the charges $P_j$ (see
Sec.~\ref{sec:quant}), $\omega$ is essentially an element of the integer
cohomology of the torus that characterizes the ensemble of cycles
wrapped by the branes.  Note that there are many different collections
of branes that will have that same $\omega$ when the harmonic
functions are one center.  For example, an arbitrary $\omega$ can be
generated by a global $U(3)$ rotation of some form characterizing
three orthogonal branes: $\omega = \sum_{i=1}^3 \tilde{P}_i dz^i\wedge
d\bar{z}^{\bar{i}}$.  As we will see below, the spacetime solution for
a collection of branes is completely characterized by $\omega$ and
this implies that in the one-center case, the {\em same} spacetime solution
is shared by many {\em different} microscopic configurations of branes.
This reflects the fact that there are many different configurations of
branes at angles that have the same asymptotically measured charges.
The no-hair theorem leads us to expect that such different sets of
branes at angles with the same asymptotic charges will have the same
spacetime solutions in the one-center case.  The way to tell these
configurations apart in the spacetime sense is to separate the
constituent branes in the noncompact space by introducing the
multi-center form of the harmonic functions.\footnote{We thank
A. Tseytlin  for discussions regarding these points.}

It is natural to define the intersection numbers
\begin{eqnarray}
C_{ij}&=&{1\over \vol(T^6)}\int_{T^6} \J\wedge \omega_i\wedge\omega_j\\
C_{ijk}&=&{1\over \vol(T^6)}\int_{T^6}
\omega_i\wedge \omega_j\wedge\omega_k.
\end{eqnarray}
The $C_{ijk}$ is proportional to the number of points at
which a T-dual collection of 4-branes intersect on the 6-torus.
This connection will be used in Sec.~\ref{sec:entropy} in computing
the BPS degeneracy of the configurations constructed in this section.
As we will see shortly, the classical solution corresponding to
2-branes at angles can be completely specified in terms of the
K\"ahler form of the asymptotic torus $k$, the 2-brane-form $\omega$,
and the number of 6-branes.

\subsection{The Classical Solution}
\label{sec:thesoln}

A classical solution corresponding to a collection of 6-branes and
2-branes at angles on a 6-torus is completely described in terms of
the metric, the dilaton, the $RR$ 3-form gauge field and the $RR$ 7-form gauge
field. To construct the solution, we need only the 2-form $k+\omega$. 
The solution in string metric is:
\begin{eqnarray}
ds^2&=&(F_2F_6)^{1/2} dx_\perp^2 + (F_2F_6)^{-1/2} \left[-dt^2+
 (h_{\mu\bar\nu} \, dz^\mu\ d\bar{z}^{\bar\nu}  +
  h_{\bar\mu\nu} \, d\bar{z}^{\bar\mu}\ dz^{\nu}) \right] \label{eq:metans}\\
A_{(3)}&=&{1\over F_2} \, dt\wedge K~~~~~~~~~~~;~~~~~~~~~~~
A_{(7)}=-{1\over F_6} \, dt\wedge {\rm dVol}
\label{eq:gaugeans}\\
e^{-2\Phi}&=&\sqrt{{F_6^3\over F_2}} \label{eq:phians}
\end{eqnarray}
where the 2-form $K$ is:
\beq 
K \equiv *{(k + \omega) \wedge (k + \omega) \over 2!}\label{eq:Kans}
\eeq
and is simply proportional to the internal K\"{a}hler metric in the
presence of 2-branes:
\beq
G = i g_{\mu\bar\nu} \, dz^{\mu} \wedge d\bar{z}^{\bar\nu} 
= {i \over \sqrt{F_2 F_6}} h_{\mu \bar\nu} \, dz^{\mu}
\wedge d\bar{z}^{\bar\nu} 
\equiv  { K \over \sqrt{F_2 F_6}} 
\eeq
The functions $F_2$ and $F_6$ have simple expressions:
\begin{eqnarray}
F_2&=&{\int_{T^6} (\J+\omega)^3\over 3!\vol(T^6)} 
= 1+\sum_i X_i +\sum_{i<j} X_iX_j \,C_{ij} +\sum_{i<j<k} X_iX_jX_k
\, C_{ijk} \label{eq:F} \\
F_6 &=& 1 + {Q_6 \over r}
\label{eq:F6}
\end{eqnarray}
Here $dx_\perp^2 = dx_7^2 + dx_8^2 + dx_9^2$ refers to the noncompact
part of the space.

The solution has a very simple structure: the 2-brane gauge field 
$A_{(3)}$ is proportional to the K\"ahler form of the compact 
manifold. Moreover, the determinant of the metric of 
the compact space is related to the dilaton through:
\be
e^{2\Phi}=\sqrt{{\rm det}~g_{\rm int}}=\sqrt{F_2\over F_6^3}
\label{eq:detgint}
\ee
as we derive in Appendix A.
These simplifying features are important for the proof of spacetime 
supersymmetry that we present in Sec.~\ref{sec:eom}.

In Sec.~\ref{sec:classical} we show that when the branes are
orthogonal to each other our construction reduces to the ``harmonic
function rule'' for orthogonally intersecting branes~\cite{harm}.  We
will also show that when only two 2-branes are present, and are
rotated at a relative $SU(2)$ angle, the solution of~\cite{mcgill1} is
reproduced.   Some of the configurations of intersecting branes
displayed here have also appeared before in~\cite{gppkt} and
subsequent publications.

To develop intuition about the solution consider
the asymptotics of the gauge fields:
\begin{eqnarray}
A_{(7)} &\stackrel{r\rightarrow\infty}{\longrightarrow}&
-dt \wedge {k^3\over 3!} ~~+~~
{Q_6 \over r} \, dt \wedge {k^3 \over 3!}  \\
A_{(3)} &\stackrel{r\rightarrow\infty}{\longrightarrow}&
dt \wedge k
~~-~~ \sum_j {P_j \over r} \, dt \wedge \omega_j
 \label{eq:3asymp}
\end{eqnarray}
(The asymptotics of $A_{(3)}$ are derived in
Appendix~\ref{app:asymp}.)  The leading terms in both of these
expressions are closed forms and do not contribute to the field
strengths $F_{(4)} = dA_{(3)}$ and $F_{(8)} = dA_{(7)}$.  The
asymptotic field strengths are therefore determined by the $1/r$
terms.  The difference in the sign of these terms between $A_{(7)}$
and $A_{(3)}$ reflects the fact explained in Sec.~\ref{sec:susy} that
only anti-6-branes can be embedded on the same torus as 2-branes at
$U(3)$ angles.  As discussed above, $k^3/3!$ is the volume form of the
asymptotic torus and and so $(-Q_6)$ is the 6-brane charge.  From
Eq.~\eq{3asymp} it is apparent that the asymptotic observer sees a
collection of 2-branes carrying charges $P_j$ and wrapped around the
cycles $\omega_j$.  The physical charges measured by this observer
relative to a canonical basis of 2-cycles at infinity will be
displayed in Sec.~\ref{sec:charges}

\subsection{Black Hole Mass and Horizon Area}

The classical solution presented in
Eqs.~\eq{metans}-\eq{phians} describes a black hole in the
noncompact four dimensions.  The mass and area of the black hole
should be computed from the four dimensional Einstein metric.  The
Einstein metric and the string metric of Eq.~\eq{metans} are
related via the four dimensional dilaton which is:
\begin{equation}
e^{-2\Phi_4} = e^{-2\Phi} \sqrt{\det g_{int}} =
\sqrt{ {F_6^3 \over F_2}}
\sqrt{ {F_2 \over F_6^3}} = 1
\end{equation} 
Here we used Eq.~\eq{detgint} for the determinant of the
metric of the compact space.  
So the four dimensional Einstein metric is
the same as the four dimensional string metric and is given by:
\begin{equation}
ds_4^2 =  (F_2F_6)^{-1/2} (-dt^2) +
          (F_2 F_6)^{1/2} (dr^2 + r^2 d\Omega^2)
\label{eq:4met}
\end{equation}
This metric describes a black hole with horizon at $r=0$.

The mass of the black hole can be read off from the behavior of the
metric as $r \rightarrow \infty$:
\begin{equation}
ds_4^2 \stackrel{r \rightarrow \infty}{\longrightarrow}  
\left(1 - { {Q_6 + \sum_j P_j} \over{2 r} } \right)
(-dt^2) +
\left(1 + { {Q_6 + \sum_j P_j} \over{2 r} } \right)
(dr^2 + r^2 d\Omega^2) .
\end{equation}
The mass $M$ is:
\begin{equation}
4G_NM = Q_6 + \sum_j P_j .
\label{eq:mass}
\end{equation}
The total mass is simply the sum of the constituent masses of the
objects in the bound state since a single 6-brane will have mass $4G_NM
= Q_6$ and each individual 2-brane has a mass $4G_NM =P_j$. Therefore
the solutions constructed here are in fact {\em marginal} bound states
in the sense that they have vanishing binding energy.

To compute the area of the horizon at $r=0$, note that the area of a
sphere of constant radius $r$ in the metric Eq.~\eq{4met} is $A =
4\pi r^2(F_2 F_6)^{1/2}$.  Now $F_6 \simeq Q_6/ r$ and:
\begin{eqnarray}
F_2 &\stackrel{r\rightarrow 0}{\longrightarrow}&  
{\int_{T^6} \omega \wedge \omega \wedge \omega \over 3!\vol(T^6)}
\nonumber \\
&=&  {1 \over r^3}
\sum_{i<j<k} P_i P_j P_k 
{\int _{T^6} \omega_i\wedge\omega_j\wedge\omega_k \over
  \vol(T^6)} = {1 \over r^3}
\sum_{i<j<k} P_i P_j P_k C_{ijk}.
\end{eqnarray}
Using this equation we find that the area of the horizon is:
\begin{equation}
A = 4\pi (Q_6\sum_{i<j<k} P_i P_j P_k C_{ijk})^{1/2}.
\label{eq:areageneral}
\end{equation}
Written this way, the formula for the area is reminiscent of the area
formulae for double extreme black holes in $N=2$ string
theory~\cite{n2area}.  Indeed, the solution constructed in this
section has relied purely on the K\"ahler nature of the compact
geometry and we are in the process of generalizing our results to
compactifications on Calabi-Yau 3-folds~\cite{us}.  In
Sec.~\ref{sec:entropy} we will write the entropy of the black hole $S=
A/4G_N$ in terms of the numbers of different kinds of branes and use
this to count the microscopic degeneracy of the corresponding D-brane
configuration.

\subsection{Canonical Charge Matrix}
\label{sec:charges}
An observer at infinity looking at the classical solutions constructed
in this section would not {\em a priori} decompose the
configuration into constituent 2-branes.   Rather, the asymptotic
observer would measure a 3-form gauge field arising from
2-brane charges associated with all 15 cycles of the 6-torus.
Considering the asymptotics of the 3-form field in Eq.~\eq{3asymp}
we are led to define the charge vector:
\begin{equation}
q_\alpha=r\int_{\Omega_\alpha} \omega=\sum_i P_i \int_{\Omega_\alpha}
\omega_i
\end{equation}
where the $\Omega_\alpha$ are a basis of 2-cycles for the 6-torus.  In
general there are 15 parameters $q_\alpha$.   However, we know from
the discussion of supersymmetry in Sec.~\ref{sec:susy} that the
$\omega_i$ are all $(1,1)$ forms relative to some choice of complex
structure.  So after a suitable choice of basis cycles, only 9 of the
$q_\alpha$ are non-vanishing.  These charges transform in the ${\bf 3}\otimes
{\bf\bar{3}}$ representation of $U(3)$ which rotates the complex
structure and it is therefore natural to assemble them into a matrix
$q_{a\bar{b}}$.   Below we will express the physical properties of the
solution - its mass and horizon area - in terms of this canonical
charge matrix.     Note that in addition to this nine parameter
matrix of charges, the solution is characterized by the 6-brane charge
giving a 10 parameter family of black holes.  In fact, configurations
of 2-branes and 6-branes on $T^6$ are characterized in general by 16
charges.   The remaining 6 parameters arise from global $SO(6)/U(3)$
rotations that deform the complex structure of the solution.   These
parameters are in one-to-one correspondence with the $(0,2)$ and $(2,0)$
cycles.  According to the analysis of Sec.~\ref{sec:susy} it is
not possible to add additional 2-branes that lie along these cycles
consistently with supersymmetry; so these charges can only be turned
on by global rotations of the entire solution.

We construct the canonical charge matrix using a basis of 2-forms
$\Omega^{a\bar b}$ that are dual to the $(1,1)$ cycles.  We can then
define a set of projection coefficients $\alpha_{ia\bar b}$ that
relate the cycles $\omega_i$ on which the 2-branes are wrapped to the
basis cycles:
\begin{equation}
\omega_i=\sum_{a \bar b} \alpha_{i a \bar b}\Omega^{a \bar b}
\end{equation}
In terms of these projection coefficients, the canonical charge matrix
is:
\begin{equation}
q_{a\bar b} = \sum_i P_i \alpha_{ia \bar b}
\end{equation}
and the 2-form $\omega$ characterizing the collection of 2-branes is
$\omega = \sum_{a \bar b} (q_{a \bar b}/r) \Omega^{a \bar b}$. Our
branes at angles construction can be used to generate an {\em
arbitrary} charge matrix.  

To express the mass and area of the black
hole in terms of the charge matrix it is convenient to choose the
${\bf 3 \otimes \bar 3}$ basis $\Omega^{a\bar b} = i\tf{a}{b}$.  In
terms of this basis the intersection form $C_{ijk}$ is:
\begin{equation}
C_{ijk} = {\int_{T^6} \omega_i \wedge \omega_j \wedge \omega_k \over
                \vol(T^6)}
 = \alpha_{ia_1\bar b_1} \,\alpha_{ja_2\bar b_2} \,\alpha_{ka_3\bar b_3}
    \epsilon^{a_1 a_2 a_3} \epsilon^{\bar{b}_1 \bar{b}_2 \bar{b}_3}        
\end{equation}
This means that:
\begin{equation}
\sum_{i<j<k} P_i P_j P_k C_{ijk} =  q_{a_1
                   \bar{b}_1} \, q_{a_2\bar{b}_2} \, q_{a_3\bar{b}_3} 
                   \, {\epsilon^{a_1 a_2 a_3} \epsilon^{\bar{b}_1
                   \bar{b}_2 \bar{b}_3} \over 3!}  = \det{q}
\end{equation}
So we can write the area of the black holes as:
\begin{equation}
A=4\pi\sqrt{Q_6\det q}
\label{eq:area2}
\end{equation}
The fact that the area formula is an invariant of global $U(3)$ rotations
is a consequence of duality.  Indeed, as will be discussed in
Sec.~\ref{sec:udual}, Eq.~\eq{area2} arises precisely as a quartic
invariant of the E(7,7)-symmetric central charge matrix of Type II
string compactified on a 6-torus. The mass of the black hole is another
invariant of the charge matrix: the trace.  To see this, note that
$Tr(q) = \sum_{ia} P_i \alpha_{i a \bar a}$.  But in the ${\bf 3 \otimes
\bar 3}$ basis,
\beql{sumofalpha}
\sum_{a} \alpha_{ia \bar a} = {\int_{T^6} \omega_i \wedge k \wedge k
\over 2! \vol(T^6) } = 1
\eeq
The last equality follows because $\omega_i$ is the volume form of a
$(1,1)$ cycle and $k$ is invariant under $U(3)$ rotations that preserve
the complex structure. (In fact, we have already used Eq.
\eq{sumofalpha} in writing Eq.~\eq{F}.)  So we find that:
\begin{equation}
4G_N M = Q_6 + \sum_j P_j = Q_6 + {\rm Tr}(q)
\end{equation}
The last equation expresses the total mass in terms of asymptotic
charges: this is the $BPS$ saturation formula.

\subsection{T-duality: 4440}
The configurations of 2-branes and 6-branes constructed above can be
converted into systems of 4-branes at angles bound to 0-branes by
T-dualizing along every cycle of the 6-torus. Below we will T-dualize
the solution of Sec.~\ref{sec:thesoln} to arrive at a solution for
4-branes at angles in the presence of a 0-brane. In this formulation,
{\em T-duality of all six directions amounts to Poincar\'{e}
duality followed by inversion of the internal metric}. As we will see, the
resulting solution is remarkably simple, compared even to that of the
2- and 6-branes.

\paragraph{Gauge Fields: }   
A 2-brane wrapped on the cycle $\omega_i$ dualizes to a 4-brane wrapped
on $\omega_{(4)i} = *\omega_i$.   Similarly, the 6-brane wrapped on
the the 6-cycle $k\wedge k \wedge k/3!$ dualizes to a 0-brane
characterized by a 0-form $\omega_0 = * (k^3/3!)$.   We conclude that
T-duality of the 6-torus acts as Poincar\'e duality on the gauge fields
of the solution.  This gives 5-form and 1-form fields in the dualized
solutions: 
\begin{eqnarray}
A_{(5)} &=& {1\over F_2} dt \wedge *K = {1\over F_2} dt \wedge 
{(k + \omega) \wedge (k + \omega) \over 2!}  \\
A_{(1)} &=& {1\over F_6} dt
\end{eqnarray}
where we used the identity $**= 1$ for $*$ acting on any
$(p,p)$ form in our conventions.

\paragraph{Metric: } The metric of the 6-torus is inverted by T-duality.
If $G = i g_{\mu\bar{\nu}} \tf{\mu}{\nu}$ is
the K\"ahler form associated with metric $g$, the K\"ahler form
associated with the inverse metric $g^{-1}$ is:\footnote{With our
conventions for the Hodge dual, $*(G \wedge G)/2 = (i/2) \left[
g_{\mu_1\bar\nu_1} \, g_{\mu_2\bar\nu_2} \right]
\epsilon^{\mu_1\mu_2}_{\hspace{0.25in}\alpha} \,
\epsilon^{\bar\nu_1\bar\nu_2}_{\hspace{0.25in}\bar\beta} \, \,
\tf{\alpha}{\beta}$.  But this precisely computes the
matrix of minors used in computing the inverse of $g$.  Dividing by
the determinant of $g$ gives the K\"ahler form of the inverse metric.}
\begin{equation}
G^{-1} \equiv *(G\wedge G)/(2!  \det{g_{\mu\bar{\nu}}})
\label{eq:inv}
\end{equation}
In our solution the K\"ahler form of the compact metric is:
\begin{equation}
G =   { K \over \sqrt{F_2 F_6}} 
= *{(k + \omega) \wedge (k + \omega) \over 2! \sqrt{F_2 F_6}} 
\end{equation}
Writing $\kappa = (k + \omega)$ 
and recalling the definition of $F_2$ in Eq.~\eq{F}, 
we see that $K/F_2 \equiv
\kappa^{-1}$ in the notation introduced in Eq.~\eq{inv}.  So
$K^{-1} \equiv \kappa / F_2 $ and the K\"ahler form of the compact
space is transformed by T-duality into:
\begin{equation}
G^\prime = G^{-1} = \kappa \sqrt{{F_6 \over F_2} }
       = (k + \omega) \sqrt{{F_6 \over F_2} }
\label{eq:invg}
\end{equation}
This also inverts the determinant of the metric of the internal space
giving $\det{g_{int}} \stackrel{T}{\rightarrow} F_6^3/F_2$.

\paragraph{Dilaton: } Under T-duality the dilaton transforms as
as $2\Phi^\prime = 2\Phi - \ln\det{g_{int}}$~\cite{giveon}.  
The determinant of the metric is computed in Appendix~\ref{app:asymp}
to be $\det g_{int} = F_2/F_6^3$.   So we find that the dilaton is
transformed by T-duality into:
\begin{equation}
e^{-2\Phi^\prime} = {e^{-2\Phi} \det g_{int}} = 
\sqrt{{F_2 \over F_6^3}}
\end{equation}

Since systems of 2-branes and 6-branes on a 6-torus are dual to
systems of 4-branes and 0-branes, we see that the most general
supersymmetric state of 4-branes on $T^6$ is constructed by wrapping
an arbitrary number of 4-branes on $(2,2)$ cycles relative to some
complex structure.  Each brane is characterized by a $(2,2)$ form
$\omega_{(4)i}$ and the collection of branes is characterized by the
form: $\omega_{(4)} = \sum_j X_j \, \omega_{(4)j}$ where $X_j =
P_j/r$ is a harmonic function on the non-compact space.  In terms of
the form $\omega$ characterizing the dual 2-branes we have:
\begin{equation}
\omega = *\omega_{(4)} = \sum_j X_j \, *\omega_{(4)j} =
\sum_j X_j \, \omega_j
\end{equation}
The intersection numbers defined in Sec.~\ref{sec:ansatz} can be
introduced again for the collection of 4-branes in terms of the
dual 2-forms.

We can now use the above discussion of T-duality to summarize the
classical solution of 4-branes at angles in the presence of 0-branes.
Define as in Sec.~\ref{sec:thesoln} the quantities:
\begin{eqnarray}
F_4(r)&=&{\int_{T^6} (\J+ *\omega_{(4)})^3\over 3!\vol(T^6)} 
= 1+\sum_i X_i +\sum_{i<j} X_iX_j \,C_{ij} +\sum_{i<j<k} X_iX_jX_k
\, C_{ijk} \label{eq:F4} \\
F_0 &=& 1 + {Q_0 \over r}
\end{eqnarray}
Relabeling $F_2$ as $F_4$ and $F_6$ as $F_0$ in the T-dual system
above, the solution is given by:
\begin{eqnarray}
ds^2&=&(F_4F_0)^{1/2} dx_\perp^2 + (F_4F_0)^{-1/2} \left[-dt^2+
 (h_{\mu\bar\nu} \, dz^\mu\ d\bar{z}^{\bar\nu}  +
  h_{\bar\mu\nu} \, d\bar{z}^{\bar\mu}\ dz^{\nu}) \right] \label{eq:met4ans}\\
G_4 &=& i g_{\mu\bar\nu} \, dz^{\mu} \wedge d\bar{z}^{\bar\nu} 
= {i  \sqrt{{1 \over F_0 F_4}}} h_{\mu \bar\nu} \, dz^{\mu}
\wedge d\bar{z}^{\nu} 
\equiv  { K_4  \sqrt{{F_0 \over F_4}}}  \\ 
K_4 
&\equiv& (k +
*\omega_{(4)})\label{eq:K4ans} \\
e^{-2\Phi}&=&\sqrt{{F_4 \over F_0^3}} \label{eq:phi4ans}\\
A_{(5)}&=&{1\over F_4} \, dt\wedge {K_4 \wedge K_4 \over
2!}~~~~~~~~~~~;~~~~~~~~~~~ 
A_{(1)}=-{1\over F_0} \, dt
\label{eq:gauge4ans}
\end{eqnarray}
As before, the metric of the compact space has been specified
in terms of its K\"ahler form $G_4$.   The four dimensional Einstein
metric is left invariant by T-duality and so these 4440 configurations
describe four-dimensional black holes just like the 2226 systems
described earlier.  The mass and the area continue to be given by
Eq.~\eq{mass} and Eq.~\eq{areageneral}.

\subsection{Mystical Comments}

The extremely simple form of the classical solution for 4- and 0-branes
is striking. The internal K\"{a}hler form is simply a sum of the
asymptotic K\"{a}hler form and the 2-form dual to the collection of
4-branes. Here we make a few remarks that hint at the underlying
geometric structure.

A 4-brane on $T^6$ has (complex) codimension one, and so it defines a
{\em divisor} of the torus. Associated with any such divisor is a line
bundle, whose first Chern class is the 2-form dual to the
divisor. Here the Chern class of the direct sum of line bundles is
making its appearance in the K\"{a}hler form, a result familiar from
symplectic geometry.  These considerations, as well as the simple
action of T-duality on the solutions point to a simple way of
including the effects of fluxes of various kinds in the brane
solutions, as we will discuss in a future article~\cite{us}.

\subsection{Example and Relation to Known Results}
\label{sec:classical}

To build intuition for our solution it is useful to work out
the details for the example with three 2-branes and one 6-brane in
Sec.~\ref{sec:2226susy}. Recall that this had three sets of branes,
rotated by different $SU(2)$ subgroups of $U(3)$.
To construct the classical solution in
Eq.~\eq{metans}-Eq.~\eq{phians} we need only compute the function
$F_2(r)$ and the K\"ahler form of the 6-torus metric.  To do this, we
begin by constructing the form $\omega$ characterizing the ensemble of
2-branes:
\begin{eqnarray}
\omega = {\textstyle \sum_j} X_j \omega_j &= & 
X_1\ i\tf{1}{1} +
X_2\ i(\cos\alpha dz^1 - \sin\alpha dz^2) \wedge
(\cos\alpha d\bar{z}^{\bar1} - \sin\alpha d\bar{z}^{\bar2})
\nonumber\\        
&&+ X_3\ i(\cos\beta dz^1 - \sin\beta dz^3) \wedge
(\cos\beta d\bar{z}^{\bar1} - \sin\beta d\bar{z}^{\bar3}) 
\label{eq:2omega}
\end{eqnarray}
Here $X_j = P_j/r$. 
We can easily compute the intersection numbers $C_{ij}$ and
$C_{ijk}$ and find that $F_2$ is given by:
\begin{eqnarray}
F_2 &= & 1 + \sum_j X_j + X_1 X_2 \, \sin^2\alpha + X_1 X_3 \,
\sin^2\beta\nonumber \\
&&+X_2 X_3 \, (\sin^2\alpha + \sin^2\beta\cos^2\alpha)
+    X_1 X_2 X_3 \, \sin^2\alpha \sin^2\beta
\label{eq:F2ex} \end{eqnarray}
The K\"ahler form of the compact space is $G = K/\sqrt{F_2 F_6}$, where
\begin{equation}
K = {*(k+\omega)^2 \over 2!} =
k + \sum_j X_j(k - \omega_j) +
\sum_{i<j} X_i X_j \, *(\omega_i \wedge \omega_j) 
\end{equation}
A small computation shows 
$*(\omega_1 \wedge \omega_2) =  \sin^2\alpha\ \Om{3}{3}$,
$*(\omega_1 \wedge \omega_3) = \sin^2\beta\ \Om{2}{2}$
and:
\begin{eqnarray}
*(\omega_2 \wedge \omega_3) &=&\left[ 
c_\alpha^2 s_\beta^2\ \Om{2}{2} + 
s_\alpha^2 c_\beta^2\ \Om{3}{3} +
s_\alpha^2 s_\beta^2\ \Om{1}{1}  \right. \\ &&+ 
\left. 
c_\beta s_\beta s_\alpha^2 \ (\Om{3}{1} + \Om{1}{3}) +
c_\alpha s_\alpha s_\beta^2 \ (\Om{2}{1} + \Om{1}{2}) +
c_\beta c_\alpha s_\beta s_\alpha \ (\Om{3}{2} + \Om{2}{3})
\right]\nonumber 
\end{eqnarray}
where we are using $\Om{a}{b}\equiv i\tf{a}{b}$, and
$s_\alpha\equiv\sin\alpha$, etc. All quantities in
Eq.~\eq{metans}-Eq.~\eq{phians} are specified in terms of $F_2$ and $K$,
and so the above formulae completely specify the spacetime solution
corresponding to the three 2-branes and the 6-brane.

\paragraph{Orthogonal Branes: } When $\alpha = \beta = \pi/2$ we have
a configuration of three orthogonal 2-branes wrapped on the (12),(34)
and (56) cycles of the torus in the presence of a 6-brane.   In that
case $F_2$  factorizes into a product of harmonic functions:
\begin{eqnarray}
F_2 &=& 1 + {\textstyle \sum_j} X_j + {\textstyle \sum_{i<j} } X_i X_j
+ X_1 X_2 X_3  = (1 + X_1) (1 + X_2) (1 + X_3) \\
e^{-2\Phi} &=& \left( F_6^3 \over (1 + X_1) (1 + X_2 ) (1 +X_3) \right)^{1/2}
\end{eqnarray}
The K\"ahler form of the 6-torus metric is given by
\begin{eqnarray}
G = {i \over \sqrt{F_6}} \left[
\tf{1}{1}\left({(1 + X_2) (1 + X_3) \over (1 + X_1) }\right)^{1/2} +
\tf{2}{2} \left({(1 + X_1) (1 + X_3) \over (1 + X_2) }\right)^{1/2}
\right. \nonumber \\
+\left. \tf{3}{3} \left({(1 + X_1) (1 + X_2) \over (1 + X_3) }\right)^{1/2}
\right]
\end{eqnarray}
and the gauge fields are:
\begin{equation}
A_{(3)} = i \, dt \wedge  \left[  {\tf{1}{1}\over (1 + X_1)} + i {\tf{2}{2}\over (1 + X_2)}
+ i {\tf{3}{3}\over (1 + X_3)} \right]  ~~~~~;~~~~~A_{(7)}={-1\over F_6} \,
dt\wedge {k^3\over 3!} 
\end{equation}
This solution coincides exactly with the ``harmonic function rule''
that governs orthogonally intersecting branes as discussed in~\cite{harm}.

\paragraph{$SU(2)$ angles: } By eliminating the third 2-brane and the
6-brane we arrive at the system discussed in~\cite{bdl,bl} -
a pair of 2-branes at a relative $SU(2)$ rotation.   Classical
solutions corresponding to such configurations were given
in~\cite{mcgill1} and discussed in~\cite{hambli}.
In this case we find: 
\begin{eqnarray}
F_2 &=& 1 + X_1 + X_2 + X_1 X_2 \sin^2\alpha \\
K &=& k + {\textstyle \sum_j} X_j (k - \omega_j) + i X_1 X_2
\sin^2\alpha \ \tf{3}{3}
\end{eqnarray}
The K\"ahler form of the metric of the compact space 
$G = K/(F_2 F_6)^{1/2}$ and the dilaton coincide with the metric
and dilaton in the solution of~\cite{mcgill1}.     To compare our 
gauge field with the one in~\cite{mcgill1} it is helpful to modify it
by adding a closed form which does not affect the field strength:
\begin{equation}
A^\prime_{(3)} = A_{(3)} + dt \wedge {k^3 \over 3!} = 
{dt \over F_2} \wedge K +   dt \wedge {k^3 \over 3!} 
\end{equation}
This gauge field coincides exactly with the one given
in~\cite{mcgill1,hambli}.\footnote{Note that the authors of~\cite{mcgill1}
are working with branes wrapped on different cycles and with charges
that are opposite to ours.  Our results for the special case of
2-branes at $SU(2)$ angles agree after some trivial relabelling of
coordinates.}

\paragraph{Mass, Area and Charges: } The mass of the example 2226
configuration is given by Eq.~\eq{mass} as $4G_N M = Q_6 + \sum_j
P_j$.   Using Eq.~\eq{areageneral} for the area of the black hole
with $C_{ijk}$ as in the computation of $F_2$ in Eq.~\eq{F2ex} we
find that:
\begin{equation}
A = 4\pi \sqrt{Q_6 (P_1 P_2 P_3) \sin^2\alpha \, \sin^2\beta}
\label{eq:areaexample}
\end{equation}
This is reminiscent of the area formula for the NS-NS black hole
generating solution presented in~\cite{ct2}.  The canonical charge
matrix defined relative to the ${\bf 3 \oplus \bar{3}}$ basis $\tf{i}{j}$ is:
\begin{equation}
q=\pmatrix{P_1 + P_2\cos^2\alpha + P_3 \cos^2\beta
& -P_2\sin\alpha\cos\alpha&
-P_3\sin\beta\cos\beta\cr
-P_2\sin\alpha\cos\alpha & P_2\sin^2\alpha &0\cr 
-P_3\sin\beta\cos\beta & 0 &P_3\sin^2\beta }
\end{equation}
It is manifest in terms of this matrix that $4GM = Q_6 +Tr(q)$ and $A =
4\pi\sqrt{Q_6\det{q}}$.

\paragraph{T-duality: } The 2226 example in this section transforms
under T-duality of the 6-torus into three 4-branes and a 0-brane at
relative angles.  The 4-form characterizing the collection of 4-branes
is $\omega_{(4)} = *\omega$.  Following
Eq.~\eq{met4ans}-Eq.~\eq{gauge4ans}, the solution is completely
characterized by $F_4$ and $K_4$.  Now $F_4 = F_2$ from the definition
of $F_4$ and the fact that $*\omega_{(4)} = **\omega = \omega$.
Furthermore $K_4 = (k + \omega)$.  It is instructive to verify that
for orthogonal branes this reproduces the ``harmonic function rule''
of~\cite{harm}.  When $\alpha =\beta = \pi/2$ we have 4-branes wrapped
on the (3456),(1256) and (1234) cycles of the 6-torus in the presence
of a 0-brane.  We find that the dilaton and  gauge field are given by:
\begin{eqnarray}
e^{-2\Phi} &=& \left({(1+X_1)(1+X_2)(1+X_2) \over F_0^3}\right)^{1/2}\\
A_{(5)} &=& dt \wedge \left[
{\Om{1}{1}\wedge\Om{2}{2}\over (1 + X_3)} +
{\Om{1}{1}\wedge\Om{3}{3}\over (1 + X_2)} +
{\Om{2}{2}\wedge\Om{3}{3}\over (1 + X_1)} 
\right]
\end{eqnarray}
The K\"ahler form of the torus is given by:
\begin{eqnarray}
G_4 = {\sqrt{F_0}} \left[
\Om{1}{1}\left({(1 + X_1) \over (1 + X_2) (1 + X_3)}\right)^{1/2} +
\Om{2}{2}\left({(1 + X_2) \over (1 + X_1) (1 + X_3) }\right)^{1/2} + 
\right. \nonumber \\
\left. +\Om{3}{3}\left({(1 + X_3) \over (1 + X_1) (1 + X_2)}\right)^{1/2}
\right]
\end{eqnarray}
This verifies the ``harmonic function rule'' for orthogonal 4-branes and
0-branes. 

\section{Spacetime Supersymmetry}
\label{sec:eom}

We have argued in Section 2 that the D-brane configurations at
arbitrary $U(3)$ angles are supersymmetric. In Section 3, we exhibited
an ansatz for the corresponding classical solution of the supergravity
equations of motion. In the present section, we prove directly that this
ansatz preserves the same supersymmetries as the corresponding
D--brane configuration. It is expected that the spacetime supersymmetry
and the Bianchi identities together imply the equations of motion,
and we verify this explicitly in Appendix~\ref{app:eom}. Thus
the Killing spinor equations are essentially the ``square root'' of
the equations of motion.

\subsection{The {\it Ansatz} and the SUSY variations}
The solutions discussed in this paper are completely determined by the
K\"ahler metric $g_{\mu\bar \nu}$ of the compact space.
We will see that the existence of
supersymmetric solutions to the supergravity equations depends
essentially only on the K\"{a}hler property of the internal metric,
and on simple relationships between that metric and the
other fields. As such, the solutions presented in this paper for
Type IIA branes on a torus are in fact special cases of a much more general
situation.  Indeed, the analysis of this section appears to carry over
almost unchanged to branes compactified on Calabi-Yau 3-folds~\cite{us}.

We begin with a solution describing 2-branes
only. In that case, the other non-vanishing fields are
given in terms of $g_{{\mu}{\bar\nu}}$ as:
\bea
g_{00}&=&-F^{-{1/2}} \label{eq:g00}\\
g_{ij}&=&\delta_{ij}F^{1/2}\\
A_3 &=& F^{-{1/2}}dt\wedge ig_{\mu{\bar\nu}}\tf{\mu}{\nu}
\label{eq:potential} \\
e^{2\Phi} &=& F^{1/2} 
\label{eq:ephi}
\eea
where
\be
\sqrt{{\rm det}~g_{\rm int}}={\rm det}~g_{{\mu}{\bar\nu}} = \sqrt{F} ~.
\label{eq:detg}
\ee 
We will use the notation $(I,J,\cdots)$ for general spacetime indices,
with $0$ as the time index, and introduce holomorphic internal indices
$(\mu,\nu\cdots)$ (and their complex conjugates), and external indices
$(i,j,\cdots)$. The
internal metric and, by implication, all the other fields, depend on the
external coordinates $x^i$ only. 
In this section 
$z^\mu=x^{2\mu-1}+ix^{2\mu}$; so the flat space metric has
$\eta_{z{\bar z}}={1\over 2}$ and 
$\eta^{z{\bar z}}=2$.\footnote{This differs from the conventions used
elsewhere in this paper. 
The $F$ in this section is denoted $F_2$ elsewhere.}
As discussed in Section~\ref{sec:ansatz}, Eqs.\eq{g00}--\eq{ephi} 
can describe an arbitrary number of 2-branes at
angles. The 
inclusion of an additional 6-brane is needed for construction of black
holes. This extension turns out to be very simple and we will consider it
in Sec.~\ref{sec:addsix}. The case of 4-branes and 0-branes, as
discussed earlier, is related by T-duality to the analysis of this section.

The supersymmetry variations of bosonic fields vanish automatically in
backgrounds that contain no fermionic fields. However, an explicit
calculation is needed to show that the variation of the fermionic fields
also vanish. In Einstein frame, these variations 
are available in \cite{dkl};   
in string frame they are:
\bea
\sqrt{2}\ \delta\lambda
&=& [{1\over 2}\partial_I \Phi~\Gamma^I\Gamma^{11}
-{3\over 16}~e^{\Phi}F_{IJ}\Gamma^{IJ}
-{i\over 192}~e^{\Phi}
F_{IJKL}
\Gamma^{IJKL}]\epsilon 
\label{eq:dilatinovar}
\\
\delta\psi_I &=& [(\partial_I+{1\over 4}
\omega_{JK,I}\Gamma^{JK})+{1\over 8}e^{\Phi}
F_{JK}
({1\over 2!}\Gamma_I^{~JK}-
\delta_I^{~J}\Gamma^K)\Gamma^{11}+\nonumber \\
&+& {i\over 8}e^{\Phi}F_{JKLM}
({1\over 4!}\Gamma_I^{~JKLM}-{1\over 3!}
\delta_I^{~J}\Gamma^{KLM})
\Gamma^{11}
]\epsilon
\label{eq:gravitinovar}
\eea The dilatino and gravitino fields are denoted $\lambda$ and
$\psi_I$, respectively, and $\omega_{JK,I}$ is the spin connection.
We are interested in configurations carrying only Ramond-Ramond
charges and so the NS 2-form $B$ has been taken to vanish.  For the
moment, we will also not include 0-, 4- or 6-branes, so the 2-form
field strength may be set to zero.  The fermionic parameters of the
two supersymmetries have been combined into a single field $\epsilon$
that is Majorana but not Weyl.

\subsection{Preliminaries}

It is worthwhile to introduce some notation and carry out various 
preliminary calculations before returning to the explicit verification 
that the supersymmetry variations indeed vanish on the stated solution.

\paragraph{Zehnbein: } In Eqs.~\eq{dilatinovar}-\eq{gravitinovar} the
supersymmetry  
variations are written using standard curved space indices.  In order
to compute the spin connection that appears in the supersymmetry
variations, it is useful to work in a local orthonormal frame.
To this end, it is convenient to introduce a
zehnbein $e^{~I}_{\hat I}$ satisfying:
\be
e^{~I}_{\hat I} e^{~J}_{\hat J} g_{IJ} = \eta_{{\hat I}{\hat J}},
\label{eq:vielb}
\ee
and their inverses $e^{~{\hat I}}_J$.  Here the indices of the local
orthonormal frame are denoted with hats.
Explicit expressions for the zehnbein can be found by solving 
Eq.~\eq{vielb}. Given the form of the metric above, we clearly have:
\bea
e^{~\hat 0}_{0} &=& \sqrt{-g_{00}} = F^{-{1/4}} \\
e^{~\hat \j}_{i} &=&  \delta_{i}^{j}F^{1/4}
\eea
In the familiar examples of parallel or orthogonally 
intersecting branes the internal metric is also diagonal,
but in the general case considered here it is not. Although
the internal components of the zehnbein can still be found, 
the expressions are unwieldy and not particularly illuminating.  
Their explicit form will not be needed.

\paragraph{Spin Connection: }The components of the spin connection
with indices along the  
transverse space are conveniently calculated from the zehnbein. They
are:
\bea
\omega_{{\hat\j}{\hat k},i}  &=& {1\over 4F}
[\delta_{ij}\partial_k-\delta_{ik}\partial_j] F 
\label{eq:ispin} \\
\omega_{{\hat\j}{\hat 0},0}  &=& -{1\over 4F^{3/2}}\partial_j F
\label{eq:tspin} 
\eea
On the other hand, the components of the spin-connection with internal 
indices are simple in their curved space form. The only non-zero 
components are:
\be
\omega_{j\nu,{\bar\mu}}
= -{1\over 2}\partial_j g_{{\bar\mu}\nu}
\label{eq:intspin}
\ee
This simple form arises because there harmonic functions appearing in
the metric depend only on coordinates transverse to the internal space.

\paragraph{Field Strength: } Next, we consider the 4-form field
strength.   The non-vanishing components 
may be computed directly from the potential (Eq.~\eq{potential}):
\be
F_{i0\mu{\bar\nu}} 
= i\partial_i (F^{-{1/2}}g_{\mu{\bar\nu}})
\ee
There will be repeated need for the trace of this expression over
the internal indices. This takes a particularly simple form:
\bea
F_{i0\mu{\bar\nu}}g^{\mu{\bar\nu}}&=&
i [3\partial_i (F^{-{1/2}})
+F^{-{1/2}}g^{\mu{\bar\nu}}\partial_i g_{\mu{\bar\nu}}) ]
\label{eq:fieldtemp} \\
&=&2i\partial_i F^{-{1/2}}
\label{eq:fieldcomp}
\eea
To obtain the last line, we have used Eq.~\eq{detg} in the form 
\be
g^{\mu{\bar\nu}}\partial_i g_{\mu{\bar\nu}}  =
\partial_i \ln{\rm det}g_{\mu{\bar\nu}} = 
{1\over 2F}\partial_i F
\label{eq:gmnident}
\ee
Note that, unlike other equations in this section, Eq.~\eq{fieldtemp}
is specific to an internal space that has three complex 
dimensions. In other cases Eq.~\eq{fieldcomp} must be established 
anew. (It is straightforward to do so for internal spaces that have two 
complex dimensions and we anticipate that other cases follow similarly.)

\paragraph{Supersymmetry: } We know that the supersymmetry is
partially broken, and so some 
constraint will be placed on the spinor $\epsilon$.
We will find solutions to Eqs. \eq{dilatinovar},\eq{gravitinovar}
provided either of the conditions
\be
\Gamma^{{\bar\mu}}\epsilon =0
\label{eq:proj1}
\ee
or
\be
\Gamma^{\mu}\epsilon =0
\ee
are satisfied. Note, for example, that the first condition is equivalent to
\beql{projnot1}
\Gamma^{\hat{\bar\mu}}\epsilon = 0
\eeq
since the
zehnbein respects the complex structure.
These conditions are entirely equivalent to the
conditions found in the D-brane analysis (see for example, 
Sec. 2.5),\footnote{That is,
Eqs. \eq{projnot1} and \eq{proj2} are equivalent to the solution in
terms of Fock space states
$|000\rangle$ and $|111\rangle$ in Sec.~\ref{sec:susy} .}
and so we are satisfied that the classical $p$-brane solutions are 
describing precisely the same D-brane configurations. 
Below, we will choose the constraint Eq.~\ref{eq:proj1}; the analysis
of the states 
satisfying the second constraint is essentially the complex conjugate
of the one presented here. 

\subsection{Proof of Supersymmetry}
With these preparations, the explicit calculations are quite simple.
In what follows we will examine the variations of the dilatino and the
external and internal components of the gravitino.

\paragraph{Dilatino: } The dilatino variation is
\be
\sqrt{2}
\delta\lambda= [{1\over 2}\partial_I \Phi \Gamma^I \Gamma^{11}
-{i\over 192}~e^{\Phi}~F_{IJKL}\Gamma^{IJKL}]\epsilon = 0
\ee
The projection conditions on the spinor Eq.~\eq{proj1} and the
identity Eq.~\eq{fieldcomp} give:
\be
{1\over 4!}F_{IJKL}\Gamma^{IJKL}\epsilon
= F_{i0\mu{\bar\nu}}\Gamma^{i0\mu{\bar\nu}}\epsilon
=-F_{i0\mu{\bar\nu}}g^{\mu{\bar\nu}}\Gamma^{i0}\epsilon
=-2i\partial_i F^{-{1/2}}\Gamma^{i0}\epsilon
\label{eq:tempf}
\ee
Also using $e^{\Phi}= F^{1/4}$ we find
\be
\Gamma^{11}\epsilon= -\Gamma^{\hat 0}\epsilon~.
\label{eq:proj2}
\ee
This condition is simply a projection on the spinor.
Note that the curved index on $\Gamma^0$ was converted to a 
flat index using $e^{~0}_{\hat 0}$. This is appropriate 
because it is  $\Gamma^{\hat 0}$ which squares to $-1$.

\paragraph{Temporal Gravitino: }The variation of the time component
of the gravitino is 
\be
\delta\psi_0 = [(\partial_0+{1\over 4}
\omega_{JK,0}\Gamma^{JK})
+ {i\over 8}e^{\Phi}F_{JKLM}
({1\over 4!}\Gamma_0^{~JKLM}-{1\over 3!}
\delta_0^{~J}\Gamma^{KLM})
\Gamma^{11}
]\epsilon =0
\ee
We will assume that the spinor is static.
Moreover, the indices on the field strength must include a $0$; so
the first of the terms with gamma-matrices vanishes, by 
antisymmetrization. Multiplication of Eq.~\eq{tempf} with $\Gamma^0$ 
gives:
\be
-{1\over 3!}F_{JKLM}\delta_0^{~J}\Gamma^{KLM}\epsilon = -2i\partial_i
F^{-{1/2}}\Gamma^i \epsilon
\ee
Using the spin connection Eq.~\eq{tspin} we recover the same 
projection condition on the spinor that was found from the 
dilatino variation (Eq.~\eq{proj2}).

\paragraph{External Gravitini: } The variations of the external
components of the gravitini are:
\be
\delta\psi_i = [(\partial_i+{1\over 4}
\omega_{jk,i}\Gamma^{jk})+{i\over 8}e^{\Phi}F_{j0\mu{\bar\nu}}
(\Gamma_i^{~j0\mu{\bar\nu}}-\delta_i^{~j}\Gamma^{0\mu{\bar\nu}})
\Gamma^{11}
]\epsilon =0
\ee
In this case we need the identity: 
\bea
F_{j0\mu{\bar\nu}}(\Gamma_i^{~j0\mu{\bar\nu}}-
\delta_i^{~j}\Gamma^{0\mu{\bar\nu}})\Gamma^{11}\epsilon &=&
-F_{j0\mu{\bar\nu}}
g^{\mu{\bar\nu}}(\Gamma_i^{~j}-\delta_i^{~j})\Gamma^0\Gamma^{11}\epsilon \\
&=& -2i\partial_j F^{-{1\over 2}}(\Gamma_i^{~j}-\delta_i^{~j})
\Gamma^{0}\Gamma^{11}\epsilon
\eea
The spin connection $\omega_{jk,i}$ (Eq.~\eq{ispin})
cancels the term proportional to $\Gamma_i^{~j}$ when the
projection Eq.~\eq{proj2} is imposed on the spinor. This leaves:
\be
(\partial_i + {1\over 8F}\partial_i F) \epsilon = 0
\ee
The solution is:
\be
\epsilon = F^{-{1/8}}\epsilon_\infty= 
e^{-\Phi/2}\epsilon_\infty~.
\ee
It is a generic property of supersymmetric solutions to supergravity
that the supersymmetry parameters depend on spacetime coordinates. The
dependence  
on the radial coordinate found here is the same as the one that occurs
in the single p-brane solutions, if both are expressed in terms of the dilaton.
A Weyl rescaled spinor can be defined that is constant
throughout spacetime.

\paragraph{Internal Gravitini: }The final equations to check are the
variations of the internal  
components of the gravitino. Given the projection \eq{proj1}, 
the components $\delta\psi_\mu$ vanish trivially. We also
need to check that 
\be
\delta\psi_{\bar\mu} = [{1\over 4}
\omega_{JK,{\bar\mu}}\Gamma^{JK}+ {i\over 8}e^{\Phi}F_{JKLM}
({1\over 4!}\Gamma_{\bar\mu}^{~JKLM}-{1\over 3!}
\delta_{\bar\mu}^{~J}\Gamma^{KLM})
\Gamma^{11}
]\epsilon =0
\ee
The identity
\bea
F_{JKLM}({1\over 4!}\Gamma_{\bar\mu}^{~JKLM}-{1\over 3!}
\delta_{\bar\mu}^{~J}\Gamma^{KLM})
\Gamma^{11}\epsilon &=& 
F_{i0{\bar\rho}\nu}(g^{\bar\rho\nu}\Gamma_{\bar\mu}^{~i0}
-2\delta_{\bar\mu}^{{\bar\rho}}\Gamma^{i0\nu})\Gamma^{11} \\
&=&2iF^{-{1\over 2}}\partial_i g_{{\bar\mu}\nu}
\Gamma^{i0\nu}\Gamma^{11}\epsilon
\eea
and the spin connection with internal indices (Eq.~\eq{intspin})
reduces the equation to the projection Eq.~\eq{proj2}, as before. 
Note that this equation relates individual components of
the gauge field and the internal metric, in contrast to previous 
conditions that only involved the determinant of the internal metric.
This tensorial structure forces the property of our solutions 
that the gauge field is proportional to the K\"ahler form of the 
compact space.

This completes the verification of spacetime supersymmetry for 
2-branes at angles.

\subsection{The addition of 6-branes}
\label{sec:addsix}
In applications to black holes, the metric describing 2-branes at
angles must be augmented with a 6-brane. This generalization is
straightforward and does not introduce issues that are not already
present for the discussion of orthogonally intersecting branes
in~\cite{harm}. The discussion will therefore be brief.

The six-branes couple to an 8-form field strength which is related
to the 2-form field strength by Hodge duality. This gives a new term 
in each supersymmetry variation.
In the presence of a 6-brane the components of the metric are
multiplied by a conformal factor. This results in an additional 
term in each of the spin connections. Similarly a term dependent on 
the 6-brane is added to the logarithm of the dilaton. In this way 
several terms are added to each spinor variation. The required
cancellations follow from the fact that the 6-brane 
is supersymmetric when there are no 2-branes present, since the
supersymmetry variations are linear in field strengths.

There is only one issue that requires detailed attention: 
the presence of the 6-brane changes the dilaton and the metric 
components in a multiplicative fashion; so it must be shown that 
such factors cancel, leaving the original supersymmetry conditions 
on the 2-brane system intact. It is straightforward to show that
the relevant precise condition is:
\be
 [e^{\Phi}e^0_{\hat 0} e^\mu_{\hat\mu} 
e^\nu_{\hat\nu}]_{\rm 6-brane~only} = F^{3/4}_6 (F^{-{1/4}}_6)^3
=1
\ee
Similarly the cancellation needed to show that the presence of the 
2-branes does not affect the 6-brane supersymmetry variations is:
\be
[e^{\Phi}e^0_{\hat 0}{\rm det}^{-1} 
g_{{\hat\mu}{\hat\nu}}]_{\rm 2-branes~only} = 
F^{1/4}~F^{1/4}~F^{-{1/2}}=1
\ee
The full 2-2-2-6 configuration is therefore supersymmetric, as advertized.

\section{Four Dimensional Black Hole Entropy}
\label{sec:entropy}
In previous sections we have constructed two T-dual 16-parameter classes
of four dimensional black holes, one containing 2-branes and 6-branes
and another containing 4-branes and 0-branes.  In this section we will
show that the entropy of these black holes can be understood
microscopically as arising from the degeneracy of the corresponding
bound state of D-branes.  In general, an extremal black hole in IIA
string theory on a 6-torus has a 56 dimensional charge vector and the
entropy of the black hole is constructed from the charges in terms of
the quartic invariant of the E(7,7) duality group. In
Sec.~\ref{sec:udual} we will discuss where our black holes belong in this
parameter space and how the representation theory of $E(7,7)$ predicts
exactly the entropy formula that we have found, convincing us that we
have correctly identified two full 16-parameter subspaces of the
spectrum of four dimensional black holes.  In Sec.~\ref{sec:quant} we
rewrite the area of the black hole solutions of this paper in terms of
quantized charges to arrive at a formula for the entropy of the black
holes that is expressed purely in terms of the number of different kinds
of branes.  In Sec.~\ref{sec:count} this representation of the entropy
is matched to the microscopic degeneracy derived by counting the number
of bound states of the corresponding D-brane configuration.

\subsection{Black Holes and U-duality}
\label{sec:udual}

According to the no-hair theorem black holes are completely
characterized by their mass, angular momentum, and $U(1)$ charges. For
extremal black holes in four dimensions the angular momentum vanishes
and the mass is given by the $BPS$ formula; so the $U(1)$ charges are
the only independent parameters. In Type II supergravity compactified on
$T^6$, the $U(1)$ charges transform in a ${\bf 56}$ dimensional
representation of the $E(7,7)$ duality 
symmetry\footnote{Some useful references for 
this subsection are~\cite{cremmer,hull,hull96}.}. There are also scalar
fields (moduli) present, but they are not independent parameters: a
given $U(1)$ charge is always multiplied by a certain combination of
moduli. It is in fact these ``dressed'' charges, {\it i.e.} charges with
moduli absorbed in them, that appear in our supergravity solutions. The
moduli parametrize the coset $E(7,7)/SU(8)$; so $SU(8)$ is the duality
group that transforms the dressed charges, but leaves the moduli
invariant.  Since the moduli parametrize inequivalent vacua, this
reduction of symmetry is simply the phenomenon of spontaneous symmetry
breaking.

The charges transform in the antisymmetric tensor representation
of $SU(8)$. They can be represented schematically in terms of 
the central charge matrix of the supersymmetry algebra:
\be
{\cal Z}_{AB}=
{1\over 4}\left( 
\begin{array}{cc}
 (Q_{R}+iP_{R})_{ab} &  {\bf R}_{ab}  \\
 -{\bf R}_{ba} & (Q_{L}+iP_{L})_{ab}   
\end{array}
\right)
\label{eqn:cencharge}
\ee 
Here $Q_{R,L}$ and $P_{R,L}$ are the NS charge vectors, written in the
spinor representation, and the $4\times 4$ matrix ${\bf R}$ 
contains the 32 RR
charges. In this paper we are interested in configurations that do not
couple to NS-fields. Note that the matrix ${\bf R}$ transforms in a
complex $({\bf 4,4})$ representation of the $SU(4)_R\times SU(4)_L$
subgroup of $SU(8)$. The geometric $SO(6)\sim SU(4)$ global rotation
group on the compact space is embedded into the $SU(4)_R\times SU(4)_L$
in such a way that ${\bf R}$ transforms under it as 
$2({\bf 4}\otimes {\bf\bar4})={\bf 1}\oplus 
{\bf 15}\oplus {\bf 15}\oplus {\bf 1}$. Here the the two ${\bf 1}$s
are clearly the 0-branes and 6-branes and the ${\bf 15}$s are the
2-branes and 4-branes.  It is convenient to  work in
an $SO(6)$ basis, in which the explicit form of the matrix ${\bf R}$ is: 
\begin{equation}
{\bf R}_{a\bar{b}}= (-Q_{6}+iQ_0)\delta_{a{\bar b}}+({1\over 2!}r_{ij}+ 
{i\over
4!}\epsilon_{ijklmn}r^{klmn})\Gamma^{ij}_{a\bar{b}} 
\label{eq:Rmatrix}
\end{equation}
where $-Q_{6}$ and $Q_0$
are the D6- and D0-brane charges, $r_{ij}$ and $r^{klmn}$ are the
projections of the D2- and D4-brane charge matrices on to the $15$
different 2-cycles and 4-cycles respectively, 
and the matrices $\Gamma^{ij}_{a\bar b}$ are generators of
$SO(6)$ in the spinor representation\footnote{The sign of $Q_6$
has been chosen so that $(-Q_6)$ is the 6-brane charge, consistently
with Sec.~\ref{sec:ansatz}.}
 The $\Gamma^{ij}_{a\bar b}$ can be
interpreted as Clebsch-Gordon coefficients that realize the equivalence
$SU(4)\simeq SO(6)$.  This decomposition is clearly consistent with the
interpretation of the charge matrix in terms of 6-branes,
2-branes, 4-branes and 0-branes.

Consider an arbitrary 2-brane charge tensor $r_{a\bar b} = r_{ij}
\Gamma^{ij}_{a\bar b}/2$.  Since $r$ is an element of the Lie
algebra of $SO(6) \sim SU(4)$ it transforms in the adjoint
representation of the group.  
From Sec.~\ref{sec:charges} we know 
that our branes at {\em relative} $U(3)$ angles can be used to
generate a family of charge matrices $q$ that fill out a $U(3)$
subalgebra of the space of $SU(4)$ Lie algebra elements comprising the
general charge matrix $r$. It is readily shown that the orbit of
a $U(3)$ subalgebra of $SU(4)$ under the action of the quotient group
$SU(4)/U(3)$ covers the entire $SU(4)$ algebra.\footnote{To show
this we use the fact that if $L$ is a $U(3)$ subalgebra, then there are no
generators $w$ of $SU(4)/U(3)$ with the property that $[w,l] \in L$
for all $l \in L$.}  The configurations of branes constructed in
this paper were wrapped on $(1,1)$ cycles relative to some complex
structure.   Since $U(3)$ is the group that preserves the complex
structure, the $SU(4)/U(3)$ global rotations generating the general
charge matrix from our configurations are simply understood as complex
structure deformations associated with the $(2,0)$ and $(0,2)$ cycles.
This shows that, after allowing for global rotations, our solutions 
realize the most general charge configuration.

We will now write the area formula for a black hole with arbitrary
charges by exploiting the fact that the Einstein metric is invariant
under duality, so that expressions for the mass and the area must be
invariant functions of the charges.  In fact the entropy $S = A/4G_N$
of a black hole has the much stronger property that it does not 
depend on moduli  and so must by expressible in terms of
the integral quantized charges as opposed to the physical ``dressed''
charges appearing in the supergravity
solutions~\cite{structure}.  This can be proven using
supersymmetry~\cite{kallosh96b}.  The entropy is therefore  
invariant under the full non-compact $E(7,7)$~\cite{kallosh96a}.
Generalizing from the simplest examples of orthogonally
intersecting branes, it follows that the area of the most general 
extremal four dimensional black hole in $N=8$ supergravity is
$A=4\pi\sqrt{J_4}$ where 
$J_4$ is the unique quartic invariant of $E(7,7)$:
\begin{equation}
J_4 = Tr ~({\cal Z}^\dagger {\cal Z})^2 
-{1\over 4}(Tr~{\cal Z}^\dagger {\cal Z})^2 
+{1\over 96}(
\epsilon_{ABCDEFGH}{\cal Z}^{AB}{\cal Z}^{CD}{\cal Z}^{EF}{\cal Z}^{GH}
+c.c )
\end{equation}
As written in Eq.~\ref{eq:Rmatrix}, the components of $R_{a\bar b}$
are not directly associated with branes wrapped on different cycles since
the $\Gamma$ matrices mix up the different components of $r_{ij}$.  
Consequently, it is easier to interpret the charges in the $SO(8)$
formalism where we rewrite the central charge matrix as:
\begin{equation}
{1\over\sqrt{2}}(x_{ij}+iy_{ij})=
-{1\over 4}{\cal Z}_{AB}(\rho_{ij})^{AB}
\label{eqn:zij}
\end{equation} 
Here $(\rho_{ij})^{AB}$ are generators of $SO(8)$. The $SO(8)$ form
of the central charge matrix is particularly useful because individual
branes correspond to specific components of the $8 \times 8$ matrices
$x_{ij}$ and $y_{ij}$.  The precise correspondence between branes
wrapped on specific cycles and particular components of the $x$ and
$y$ matrices can be found by explicitly writing out the transformation
into $SO(8)$ basis.  In the case of our $U(3)$ family of D2-brane
configurations it is convenient to introduce complexified coordinates
to describe the first six of the eight $SO(8)$ indices, while keeping
the indices $7$ and $8$ unchanged. In this notation it can be shown
that for our configurations the non-vanishing components of the
central charge
matrix are $y_{a{\bar b}}= {1\over \sqrt{2}}q_{a{\bar b}}$ and 
$y_{78}=-{1\over \sqrt{2}}Q_6$
where $q_{a \bar b}$ are the canonical charges defined in
Sec.~\ref{sec:charges}.
 
The fact that in our solution the $x_{ij}$ vanish for
some choice of complex structure reflects a simplification that is 
not generic\footnote{Acting on all the branes with 
a global $SU(4)/U(3)$ rotation that does not respect the 
complex structure would turn on some $x_{ij}$.}. 
For the $SU(8)$ form of the charge matrix in Eq.~\ref{eq:Rmatrix} 
the analogous statement is that the central charge matrix can be 
chosen real.  In fact, in $N=2$ and $N=4$ supergravity the 
central charge can always be chosen real after applying suitable 
dualities, because in these cases the compact part of the duality 
group is $U(N)$. On the other hand, for $N=8$ supergravity the 
compact part of the duality
group is $SU(8)$, as we have seen, so in general the central charge
matrix has an invariant phase~\cite{ct2,hull96}.  
It follows that there are configurations in which the central 
charge cannot be chosen to be real.
Although very general, our configurations do not capture this feature
of $N=8$ supergravity.

We can now calculate from group theory the area of a black hole
carrying general 2-brane and 6-brane charges.  In the $SO(8)$
formalism the quartic invariant is:
\begin{equation}
-J_4=
x^{ij}y_{jk}x^{kl}y_{li}-{1\over 4}x^{ij}y_{ij}x^{kl}y_{kl}
+{1\over 96}\epsilon_{ijklmnop}(x^{ij}x^{kl}x^{mn}x^{op}
+y^{ij}y^{kl}y^{mn}y^{op})
\end{equation}
For our configurations only the last term of this expression
in nonvanishing:
\begin{equation}
J_4 = Q_6~{\rm det}~q
\end{equation}
This gives a black hole area of $A=4\pi\sqrt{Q_6\det{q}}$ in agreement
with the explicit calculation of Sec.~\ref{sec:charges}.  Note that
the derivation of the entropy given in this section is independent of
the explicit solution: it relies only on supersymmetry and duality
invariance.

\subsection{Quantized Charges}
\label{sec:quant}
The first step towards a microscopic understanding of the
black hole entropy is to express it in terms of the quantized charges.  
We need to understand how the physical charges $P_i$ and $Q_6$
appearing in the harmonic functions governing the solution are related
to the numbers of branes wrapped on different cycles.  As before
we will assume that the asymptotic torus is square and has
unit moduli as in the solutions presented in Sec.~\ref{sec:ansatz}.
We will work with the solution containing 2-branes and 6-branes - the
discussion of the solution containing 4-branes and 0-branes proceeds
analogously. Recall that we defined the canonical charges $q_{ja\bar
b}$ induced on the (1,1)-cycles of the torus by a given 2-brane wrapped on
the cycle $\omega_j$ via:
\begin{equation}
P_j \omega_j = \sum_{a\bar{b}} P_j \alpha_{ja \bar{b}} \, \Om{a}{b} \equiv  
\sum_{a\bar{b}} q_{ja \bar{b}} \, i\tf{a}{b} 
\end{equation}
Adapting the formulae in the Appendix of~\cite{vfscatter} for branes
compactified on $T^6$ it is easy to show the quantization condition for
the canonical 2-brane charges and the 6-brane charge:
\begin{equation}
q_{ja\bar b} = {l_s g \over 4\pi} n_{ja\bar b}~~~~~~~~;~~~~~~~~
Q_6 = {l_s g \over 4\pi} N_6
\end{equation}
Here $n_{ja\bar b}$ counts the number of times the $j$th 2-brane wraps
the $(a\bar b)$ cycle and $N_6$ counts the number of times the 6-brane
wraps the entire torus. 
The string length is $l_s=2\pi\sqrt{\alpha^\prime}$.
This quantization condition leads us to write:
\begin{equation}
P_j \omega_j = {l_s g \over 4\pi} \sum_{a\bar{b}} n_{ja\bar{b}}\Om{a}{b} 
\equiv {l_s g \over 4\pi} M_j \sum_{a\bar{b}} m_{ja\bar{b}}\Om{a}{b}
\equiv {l_s g \over 4\pi} M_j v_j
\end{equation}
where we have defined $M_j$ to be the greatest common factor of
the $n_{ja\bar b}$.   Now $M_j$ is the
integral number of branes wrapping the cycle $\omega_j$ and $v_j$ is
the element of the {\it integral} cohomology of the torus that
characterizes the cycle.  (Recall that $\omega_j$ was a {\it volume}
form and as such not a member of the integral cohomology.)  We then
define:
\begin{equation}
N_{ijk} = {1 \over \vol(T^6)} \int_{T^6} v_i \wedge v_j \wedge v_k
\end{equation}
Recall that T-duality of the 6-torus converts the 2-branes into
4-branes wrapped on the cycles $*\omega_i$.  The forms $*v_i$
characterize these dual 4-branes in integral cohomology.  Three
4-branes on a 6-torus generically intersect at a point and the 
$N_{ijk}$ are integers counting the number of intersection points.
Using the quantized charges we find that:
\begin{equation}
\sum_{i<j<k} P_i P_j P_k C_{ijk} =
\left({l_s g \over 4\pi} \right)^3 
\sum_{i<j<k} (M_i M_j M_k) N_{ijk}
\end{equation}
Putting this together with the 6-brane quantization condition we can
rewrite the area formula as:
\begin{equation}
A = 4\pi \, \left({l_s g \over 4\pi}\right)^2 \,
\sqrt{N_6 \sum_{i<j<k} ( M_i M_j M_k) N_{ijk} }
\end{equation}
Finally, the entropy of the black hole is given by $S = A/4 G_N$ and
the Newton coupling constant is $G_N = g^2 \alpha'/8 = g^2
l_s^2/32\pi^2$.   This gives:
\begin{equation}
S = 2\pi \sqrt{N_6 \sum_{i<j<k} (M_i M_j M_k) N_{ijk}}
\label{eq:entropy}
\end{equation}
In the next section we will explain the origin of this entropy
microscopically in terms of the degeneracy of the corresponding bound
state of D-branes.

\paragraph{Example: } To understand these results intuitively it is
instructive to explicitly evaluate the entropy in terms of quantized
charges in the example of Sec.~\ref{sec:classical}.  We will work in
the T-dual picture where there are three 4-branes and examine the
meaning of the quantization conditions to show how the intersection
number of branes arises in the entropy formula.  The first brane is
wrapped on the coordinates $(y_3,y_4,y_5,y_6)$, the second is on
$((s_\alpha y_1 + c_\alpha y_3), (s_\alpha y_2 + c_\alpha y_4), y_5,
y_6)$ and the third is on $((s_\alpha y_1 + c_\alpha y_5), (s_\alpha
y_2 + c_\alpha y_6), y_2, y_3)$.  (We have adopted the notation of
Sec.~\ref{sec:classical} and  taken $\omega_{(4)i} = *\omega_i$ where the
$\omega_i$ characterize the 2-branes in Eq.~\eq{2omega}).  
The area of the resulting black hole, as evaluated in Eq.~\eq{areaexample},
is:
\begin{equation}
A = 4\pi \sqrt{Q_0 P_1 P_2 P_3 \sin^2\alpha \sin^2\beta}
\label{eq:area4ex}
\end{equation}
(We have relabelled $Q_6$ as $Q_0$ after T-duality.)  Since the
4-branes must wrap a finite number of times around the torus, the
angles $\alpha$ and $\beta$ are quantized as $\tan_\alpha =
q_\alpha/p_\alpha$ and $\tan_\beta = q_\beta/p_\beta$ where each $q,
p$ pair are relatively prime integers. Projecting the quantization 
conditions discussed above on to
the cycles defined by the branes we write the $P_i$ as:
\begin{equation}
P_i = {l_s g \over 4\pi} M_i A_i
\end{equation}
where $M_i$ is the wrapping number of the ith brane and $A_i$ is its 
dimensionless area ($A_1 = 1$,
$A_2 = p_\alpha^2 + q_\alpha^2$, and $A_3 = p_\beta^2 + q_\beta^2$).
This makes good sense since each brane has a fixed charge density and
so the total physical charge of a brane should be proportional to its
area.  So we see that the quantization condition for the canonical
charges $q_{ia \bar b}$ produces the correct quantization of the
charges of the angled branes. Inserting these quantized charges into
the area Eq.~\eq{area4ex} and dividing by $4G_N$ gives the
entropy:
\begin{equation}
S = 2\pi \sqrt{N_0 (M_1 M_2 M_3) q_\alpha^2 q_\beta^2}
\end{equation}
We now want to argue that $q_\alpha^2 q_\beta^2$ counts the number of
intersection points of the three 4-branes on the 6-torus. Consider 
a string wound on the (0,1) cycle of a 2-torus and another string on 
a (q,p) cycle.  It is clear that the strings
intersect q times.  In our case we have one 4-brane on the (3456)
cycle and another one on the ([13][24](56)) cycle.  (The square
brackets indicate angling on the corresponding torus.)  The first
brane is on a (0,1) cycle on the (13) and (24) 2-tori.  The second
brane is on a $(q_\alpha,p_\alpha)$ cycle on both these tori.  So it
is clear that the two 4-branes intersect in $q_\alpha^2$ places on the
(1234) torus and each intersection has two-dimensional extent along
the (56) cycle.   Each of these intersection manifolds lying on the (56)
cycle is intersected in $q_\beta^2$ places by the third 4-brane
leading to $q_\alpha^2 q_\beta^2$ mutual intersections of the three
4-branes.   So the factor of $\sin^2\alpha \sin^2\beta$ in
Eq.~\eq{area4ex}, when 
multiplied by the factors of area in the quantization condition for
angled branes, precisely reproduces the intersection number of the
three 4-branes as discussed more abstractly above.

\subsection{Counting the States of The Black Hole}
\label{sec:count}
Counting the states of the black hole is easiest in the picture with
4-branes and 0-branes where the entropy formula is $S= 2\pi \sqrt{N_0
\sum_{i<j<k} (M_i M_j M_k) N_{ijk}}$.  In this case the analysis is
exactly parallel to the one described in~\cite{vf,kt}.  More recently,
a similar discussion has appeared for certain black holes in $N=2$
string theory in~\cite{juan1,behrndt2}.  The counting is aided by the
M-theory perspective where 0-brane charge arises as momentum in the
11th dimension and 4-branes are the dimensional reduction of M-theory
5-branes wrapped on the 11th dimension.  Three intersecting 4-branes
arise in 11 dimensions as three 5-branes intersecting along a line and
the 0-branes that we are interested in arise as momentum along one
direction of this line.

\paragraph{Large $N_0$:}  The argument is
simplest when $N_0$, the number of 0-branes, greatly exceeds the number
of 4-brane intersections.  The leading contribution to the black hole
entropy arises in this case from the number of different ways in which
a total 0-brane charge of $N_0$ can be distributed between the 4-brane
intersections.  Since the momentum in the 11th dimension can come in
integral multiples, the charge of 0-branes bound to the mutual
intersection of three 4-branes can also come in integral multiples.
We now {\it assume} that the  0-branes bound to the 4-branes, or
momentum modes along the intersection string along the 11th dimension
give rise to states that appear as $B$ bosonic and $F$ fermionic 
species with an associated central charge of 
$c = B + (1/2) F= 6 $.  (See
Sec.~\ref{sec:effstring} for the origin of this assumption in an
effective string description of the black hole entropy.)  Since there
are a total of $N_{int} = \sum_{i<j<k} (M_i M_j M_k) N_{ijk}$
intersection points of 4-branes, the problem is very simply to count
the number of ways of distributing $B$ bosonic and $F$ fermion
0-branes that come in integrally charged varieties amongst $N_{int}$
intersections.  From the 11-dimensional perspective, we want to
distribute a total momentum of $N_0$ carried by $B$ bosonic and $F$
fermionic modes amongst $N_{int}$
strings.  As discussed in~\cite{vf,kt}, both of these problems are
identical to the computation of the density of states at level $N_0$
of a string with central charge $c_{\rm eff} = c N_{int} = 6 N_{int}$.
For large $N_0$, the level density is $d(N_0) = \exp{2\pi\sqrt{N_0
c/6}}$ giving an entropy of~\cite{gsw}:
\begin{equation}
S= \ln{d(N_0)} = 2\pi\sqrt{N_0 \sum_{i,j,k} (M_i M_j M_k) N_{ijk}}
\label{eq:rep1}
\end{equation}
This exactly reproduces Eq.~\eq{entropy}, the entropy formula for the
black hole!   

\paragraph{General $N_0$: }  The above counting of states only works
in the limit when $N_0$ greatly exceeds the number of 4-brane
intersections,  because that is the regime of validity of the
asymptotic level density 
formula that gives rise to the entropy. In the general case
where the number of 0-branes is comparable to the number of 4-brane 
intersections, the leading contribution to the black hole entropy comes
from states in which the 4-brane charges arise from {\em
multiply-wrapped} branes rather than from multiple {\em
singly-wrapped} branes.  From the M-theory perspective, the multiple
wrappings of the 5-branes increase the length of the effective string
along which the momentum propagates, changing the quantization
condition for the momentum and hence the 0-brane charge.  As discussed
in~\cite{mathur96,susskind96}, this corrects the level density
asymptotics  in exactly the right way to make the state counting
discussed above applicable to the large charge classical black hole
regime.   The application of~\cite{mathur96,susskind96} to precisely
this context of 4-branes and 0-branes is discussed in~\cite{vf,kt}.

\subsection{Effective String Description}
\label{sec:effstring}
The discussion in Sec.~\ref{sec:count} relied on the assumption that
the momentum modes on the mutual intersection line of three M-theory
5-branes have $B$ bosonic and $F$ fermionic degrees of
freedom so that $c = B + (1/2)F = 6$.  Some arguments in favor of
this assumption were presented in~\cite{vf,kt}.  Certainly the
assumption seems to succeed in accounting for the entropy of black
holes in a wide variety of cases in Type II strings compactified on
both tori~\cite{vf,kt} and on Calabi-Yau
3-folds~\cite{juan1,behrndt2}.  This suggests that we should turn the
argument around and use the matching with the entropy of black holes
to derive the physics of the intersection manifold of 5-branes in
M-theory.

Three 5-branes intersecting along a line in 11 dimensions have a (0,4)
supersymmetry on that line.  This means that only left-moving momentum
can be added to the string without breaking supersymmetry as indeed we
found in Sec.~\ref{sec:susy} after T-dualizing the resulting 0-branes
in 10 dimensions into 6-branes.  While the mutual intersection line of
the 5-branes is wrapped along the 11th dimension, the other
dimensions of the 5-branes are wrapped on a $T^6$ in our solution.
Consider making the $T^6$ small while leaving the 11th dimension
large.  The resulting effective string should be described by a
$(0,4)$ supersymmetric sigma model on the orbifold target:
\begin{equation}
{\cal M} = {(T^6)^{M_i M_j M_k} \over S(M_i M_j M_k)}
\label{eq:orbifold}
\end{equation}
where the $M_i$ are the quantized charges of the three 5-branes and we
orbifold by the symmetric group $S(M_i M_j M_k)$ to account for
symmetry under exchange of the 5-branes.\footnote{Note that one might
have naively supposed that that appropriate orbifold group would be
$S(M_i)S(M_j)S(M_k)$ which would account for the exchange symmetry of
each kind of 5-brane.  However, the analysis of~\cite{sv1,dmvv}
indicates that the appropriate group is $S(M_1 M_2 M_3)$. Indeed, this
is the orbifold that is consistent with T-duality. }  The asymptotic
degeneracy of BPS states of this effective string will reproduce the
entropy formula Eq.~\eq{entropy} for the case where only three
5-branes are present, following the work of~\cite{dvv1,dvv3,dmvv}.

In our case, each triplet of 5-branes intersects in $N_{ijk}$
locations giving rise to $N_{ijk}$ effective strings.  So from the
M-theory perspective, our solutions are described in the small torus
limit by $N_{tot} = \sum_{i<j<k} N_{ijk}$ effective strings, each
propagating on an orbifold like Eq.~\eq{orbifold}.  The appropriate
total effective conformal field theories have central charges that are 
the sum of contributions from many effective strings, and the resulting
degeneracy exactly matches our Eq.~\eq{entropy}.

\section{Conclusion}
In this paper we have shown that the most general supersymmetric state
of 2-branes on a 6-torus is constructed using an arbitrary number of
branes at relative $U(3)$ angles, acted upon by global $SO(6)/U(3)$
rotations.  After addition of 6-branes, these configurations account
for a 16-parameter subspace of the spectrum of BPS states of IIA
string theory on $T^6$.  T-duality of the torus converts these states
into the most general BPS configurations of 4-branes and 0-branes on a
6-torus.  We have constructed the corresponding solutions to the
supergravity equations and verified explicitly that they solve the
Killing spinor equations.  The spacetime solutions are remarkably
simple when expressed in terms of the complex geometry of the compact
space.  Very little in our analysis has relied on toroidal structure
of the compact space - the relevant properties are that the compact
space is K\"ahler and the gauge fields are proportional to the
K\"ahler form.  In fact, we expect our results to go through with only
slight modifications for compactifications on Calabi-Yau 3-folds,
which we are in the process of verifying~\cite{us}. The geometric
structure of our solutions suggests natural generalizations that would
provide the spacetime solutions for arbitrary BPS bound states of Type
II solitons.  We are in the process of investigating these
generalizations as well.  The configurations constructed in this paper
can be interpreted as black holes in four dimensions.  We computed the
thermodynamic entropies of these black holes and showed that they 
can be interpreted
microscopically in terms of the bound state degeneracy of the
corresponding collection of D-branes.  Our calculations agree
beautifully with the analysis of four dimensional black hole entropy
in terms of the quartic invariant of the $E(7,7)$ duality group.

\vspace{0.2in} {\bf Acknowledgments:} We would like to thank C. Callan
and M. Cveti\v{c} for numerous interesting discussions and
G. Papadopoulos and E. Sharpe for helpful comments.  We have also had
useful correspondence with A. Tseytlin.  V.B. was supported partly by
DOE and NSF grants DE-FG02-91-ER40671 and NSF-Phy-91-18167 and also by
the Harvard Society of Fellows.  F.L. was supported in part by DOE
grant AC02-76-ERO-3071.

\appendix
\section{Useful Properties of the Solutions}
\label{app:asymp}

In this appendix we will derive the asymptotics of the 3-form
gauge field and then compute the determinant of the metric of internal
6-torus.  Throughout this paper we are using the following definition
of the Hodge dual of $(p,q)$ forms:
\begin{eqnarray}
*(dz^{\mu_1} \wedge \cdots dz^{\mu_p} \wedge 
d\bar{z}^{\nu_1} \wedge \cdots d\bar{z}^{\nu_q}) 
= & i^{p+q+1} {1  \over (3-p)! \, (3-q)!}
\epsilon^{\mu_1\cdots\mu_p}_{\hspace{0.3in}\alpha_1\cdots\alpha_{3-p}}
\epsilon^{\nu_1\cdots\nu_q}_{\hspace{0.3in}\beta_1 \cdots \beta_{3-q}} 
\times \nonumber \\ 
&dz^{\alpha_1}  \wedge \cdots dz^{\alpha_{3-p}} 
\wedge 
d\bar{z}^{\beta_1} \wedge \cdots d\bar{z}^{\beta_{3-q}}
\end{eqnarray}
where $\epsilon_{123} = 1$ and the indices are raised and lowered
using the flat metric of the asymptotic torus.

\paragraph{3-form Asymptotics: } The 3-form gauge field is given by
$A_{(3)} = (1/F_2) dt \wedge K $.  We want to extract the leading $r$
dependence at large $r$.  As $r \rightarrow \infty$, $1/F_2
\rightarrow 1 - \sum_j X_j$.  We also 
have $K= *(k + \omega)^2/2 
\sim k + 2 \sum_j X_j *(k \wedge \omega_j)$ for large $r$.  
In the conventions of Sec.~\ref{sec:susy} and Sec.~\ref{sec:ansatz}, 
$\omega_j$ is a $U(3)$ rotation of the $(1,1)$ form $dz^1 \wedge
d\bar{z}^1$.  The K\"ahler form $k$ of the asymptotic torus is
invariant under such $U(3)$ rotations that preserve the complex structure,
and it is easy to use this and the definition of the Hodge dual
to show that:
\begin{equation}
*(k \wedge \omega_j) = (k - \omega_j)
\end{equation}
This gives the asymptotics:
\begin{equation}
A_{(3)} \stackrel{r \rightarrow \infty}{\longrightarrow}
dt \wedge k
~-~ \sum_j {P_j \over r} \, (dt \wedge \omega_j)
\end{equation}
where we used the definition $X_j = P_j / r$ for the harmonic function
associated with the jth brane.

\paragraph{Determinant of 6-torus metric: }  The metric of the
compact space $g_{int}$ is most readily specified in terms of its
K\"ahler form:
\begin{equation}
G = {1 \over \sqrt{F_2 F_6} }  \, K 
\end{equation}
where $K = *\kappa^2/2$ and $F_2 = \int_{T^6} \kappa \wedge \kappa
\wedge \kappa / 3!\,\vol(T^6)$ with $\kappa = (k + \omega)$.  Writing 
$G = i g_{\mu\bar\nu} dz^\mu \wedge
d\bar{z}^\nu$, while the metric of the compact space is $ds^2 =
g_{\mu\bar\nu} dz^\mu d\bar{z}^\nu + g_{\bar\mu\nu} d\bar{z}^\mu
dz^\nu $, we find:
\begin{equation}
\sqrt{\det{g_{int}}} =  {1 \over 3! \, \vol(T^6)}
                      \int_{T^6} G \wedge G \wedge G  
= {1 \over    3! \, \vol(T^6) \,     {\sqrt{F_2^3 F_6^3}}}
\int_{T^6} K \wedge K \wedge K
\label{eq:det}
\end{equation}
To compute $K \wedge K \wedge K$ it is helpful to write
$\kappa = (k + \omega) = i f_{\mu \bar\nu} dz^\mu \wedge d\bar{z}^\nu$
so that $\kappa \wedge \kappa \wedge \kappa/3! = \det(f_{\mu\bar\nu}) \,
dV$ where $dV$ is the volume form of the torus.  Using
Eq.~\eq{F} this means that $\det(f) = F_2$.  In terms of $h$, the
form $K$ is given by:
\begin{equation}
K = *{(\kappa^2)\over 2} = 
{i \over 2} \left[ f_{\mu_1\bar\nu_1} \, f_{\mu_2\bar\nu_2}
\right] \epsilon^{\mu_1\mu_2}_{\hspace{0.25in}\alpha} \,
\epsilon^{\bar\nu_1\bar\nu_2}_{\hspace{0.25in}\bar\beta} \,
\, dz^\alpha \wedge d\bar{z}^\beta
\end{equation}
After some manipulation of the $\epsilon$ tensors and taking into account
the symmetries of the metric, we find that:
\begin{equation}
{K \wedge K \wedge K \over 3!} = dV \, \det(f)^2 = dV \, F^2_2
\end{equation}
Using this is Eq.~\eq{det} gives the result:
\begin{equation}
\sqrt{\det{g_{int}}} = {1 \over \sqrt{F_2^3  F_6^3} } F_2^2 = 
\sqrt{ F_2 \over F_6^3}
\end{equation}

\section{The equations of motion}
\label{app:eom}

The purpose of this appendix is to show explicitly that the classical 
configurations considered in this paper do indeed satisfy the equations 
of motion. This computation is straightforward in principle but,
unless properly organized, it can be prohibitively tedious. We restrict 
the attention to the case of D2-branes at angles, because the inclusion 
of D6-branes involves no essential new features.

In Sec.~\ref{sec:eom} we showed that the classical configurations
Eq.~\ref{eq:g00}--\ref{eq:ephi} are supersymmetric in spacetime. 
These supersymmetric configurations are completely characterized by 
the K\"{a}hler metric $g_{\mu\bar{\nu}}$ of the compact space. It is 
important to note that the dependence of $g_{\mu\bar{\nu}}$ on the 
uncompactified 
coordinates is not restricted by supersymmetry. However, our {\it ansatz} 
Eq.~\ref{eq:metans}-\ref{eq:F6} is more restrictive: the 
$g_{\mu\bar{\nu}}$ are given in terms of the $X_i$ that in turn are 
{\it harmonic functions} of the external coordinates. We can use 
Eq.~\ref{eq:invg} to express this additional requirement conveniently
as:
\be
\partial^2 (F^{1\over 2}g^{\mu\bar{\nu}}) = 0
\label{eq:intharm}
\ee
It is easy to use the Bianchi identities to show that this harmonic 
function property is indeed a necessary condition to satisfy the equation 
of motion. In this section we will verify that this condition is also 
sufficient.

Note that in Eq.~\ref{eq:intharm}, and repeatedly in the following, 
the $\partial_i$ denote derivatives with respect to the 
{\it Cartesian coordinates} in the noncompact space, and the  
index contraction implied in the notation $\partial^2\equiv 
\partial_i \partial^i$ is with respect to the Cartesian flat metric 
in three dimensions, {\it i.e.} $\delta_i^j$. 

\subsection{Preliminaries}
We use the notation explained in the beginning of Sec.\ref{sec:eom}. 
Our strategy is to use the {\it ansatz} Eq.~\ref{eq:g00}--\ref{eq:ephi} 
to express all quantities in terms of the functions $g_{\mu\bar{\nu}}$ 
and $F$. 
Subsequently the harmonic function property Eq.~\ref{eq:intharm} 
is employed to simplify the expressions.

\paragraph{Equations of motion:}
We write the part of the Lagrangian that is needed for the present
purposes as:
\be
{\cal L} = \sqrt{-g} [ e^{-2\Phi} ( R + 4 (\nabla\Phi)^2) - 
{1\over 48}F^2_4 ]
\ee
where $F_4$ is the 4-form field strength that couples to the D2-branes.
The corresponding equations of motion can be written:
\bea
R &=& -4\nabla^2 \Phi + 4 (\nabla\Phi)^2 
\label{eq:dilaton} \\
R_{IJ} &=& - 2\nabla_I\nabla_J \Phi
+{1\over 12}e^{2\Phi}(F_I^{~KLM}F_{JKLM}-
{1\over 8}g_{IJ}F^{KLMN} F_{KLMN})
\label{eq:gravity}\\
0 &=& {1\over\sqrt{-g}}\partial_I (\sqrt{-g} F^{IJKL}) 
\label{eq:gauge}
\eea
We first consider the left hand side of these equations.

\paragraph{Curvature:}
The nonvanishing components of the Christoffel symbols are:
\bea
\Gamma_{ijk} &=& {1\over 4} F^{-{1\over 2}}(\delta_{ik}\partial_j F
+\delta_{ij}\partial_k F-\delta_{jk}\partial_i F) \\
\Gamma_{i00} &=& -{1\over 4}F^{-{3\over 2}} \partial_i F 
= -\Gamma_{0i0} \\
\Gamma_{i\mu\bar{\nu}} &=& -{1\over 2}\partial_i g_{\mu\bar{\nu}}
= - \Gamma_{\mu i\bar{\nu}} = -\Gamma_{\bar{\nu}i\mu}
\eea
We then use:
\be
R_{IJ} = \partial_K \Gamma^K_{IJ} - \partial_I \Gamma^K_{KJ}
+ \Gamma^K_{IJ}\Gamma^L_{KL} - \Gamma^K_{LJ}\Gamma^L_{KI}
\ee
to find the nonvanishing components of the Ricci tensor:
\bea
R_{00} &=& -{1\over 4}F^{-2}\partial^2 F + 
{1\over 8}F^{-3}(\partial F)^2 
\label{eq:R00}\\
R_{ij} &=& -{1\over 2}F^{-1}\partial_i\partial_j F - 
{1\over 4}\delta_{ij}F^{-1}\partial^2 F + {5\over 8}F^{-2}
\partial_i F\partial_j F+{1\over 8}\delta_{ij}F^{-2}(\partial F)^2+ 
\label{eq:Rij} \nonumber \\
&+&{1\over 2}\partial_i g^{\mu\bar{\nu}}\partial_j g_{\mu\bar{\nu}} \\
R_{\mu\bar{\nu}} &=&
-{1\over 2}F^{-{1\over 2}}\partial^2 g_{\mu\bar{\nu}}
-{1\over 4}F^{-{3\over 2}}\partial F \partial g_{\mu\bar{\nu}}
+{1\over 2}F^{-{1\over 2}}\partial g_{\mu\bar{\rho}}
g^{\bar{\rho}\lambda}\partial g_{\lambda\bar{\nu}}
\eea
In the evaluation we used:
\be
\Gamma^I_{Ik} = {1\over 2g}\partial_k g = F^{-1}\partial_k F
\ee
The harmonic function property Eq.~\ref{eq:intharm} can be employed to 
show:
\bea
\partial^2 g_{\mu\bar{\nu}} &=&
[{1\over 2}F^{-1}\partial^2 F-
{1\over 4}F^{-2}(\partial F)^2] g_{\mu\bar{\nu}}
-F^{-1}\partial F\partial g_{\mu\bar{\nu}} 
+ 2\partial g_{\mu\bar{\rho}}
g^{\bar{\rho}\lambda}\partial g_{\lambda\bar{\nu}}\\
\partial g^{\mu\bar{\nu}}\partial g_{\mu\bar{\nu}}  
&=& F^{-1}\partial^2 F - {3\over 4} F^{-2}(\partial F)^2 
\label{eq:iden2}
\eea
The first relation allows us to write the Ricci tensor with indices 
in the compact directions as:
\be
R_{\mu\bar{\nu}} =
[-{1\over 4}F^{-{3\over 2}}\partial^2 F +{1\over 8}F^{-{5\over 2}}
(\partial F)^2]g_{\mu\bar{\nu}}
+{1\over 4}F^{-{3\over 2}}\partial F \partial g_{\mu\bar{\nu}}
-{1\over 2}F^{-{1\over 2}}\partial g_{\mu\bar{\rho}}
g^{\bar{\rho}\lambda}\partial g_{\lambda\bar{\nu}}
\label{eq:Rmn}
\ee
and the second relation is needed to find the Ricci scalar:
\bea
R &=& g^{00}R_{00} + g^{ij}R_{ij} + 2g^{\mu\bar{\nu}}R_{\mu\bar{\nu}}\\
&=& -F^{-{3\over 2}}\partial^2 F+{3\over 4}F^{-{5\over 2}}(\partial F)^2
\label{eq:ricscalar}
\eea

\subsection{The equations of motion}
At this point we have evaluated the left hand side of the equations
of motion Eq.~\ref{eq:dilaton}-\ref{eq:gauge}, {\it i.e.} the
gravity contribution. We now proceed to consider the right hand side, 
{\it i.e.} the matter contribution.

\paragraph{The gauge field equation}
The simplest and most instructive of the equations of motion is
Eq.~\ref{eq:gauge} for the 4-form field strength. The only
component that is nontrivial is:
\be
{1\over\sqrt{-g}}\partial_I (\sqrt{-g} F^{It\mu\bar{\nu}})
= F^{-1}\partial_i (g^{ij}
F g^{\bar{\nu}\nu}g^{\bar{\mu}\mu}
\partial_j  F^{-{1\over 2}}g_{\bar{\mu}\nu})
=-F^{-1}\partial^2 F^{1\over 2}g^{\mu\bar{\nu}}
= 0
\ee
as it should be.
Note that the final step requires exactly the harmonic function
property Eq.~\ref{eq:intharm}. This calculation therefore
demonstrates the necessity of this condition.
In fact, it is precisely the Bianchi identity that is verified
in this step.
 
\paragraph{The dilaton equation:}
The {\it ansatz} Eq.~\ref{eq:metans}-\ref{eq:F6} gives:
\bea
\nabla^2 \Phi &=& {1\over 4}F^{-{3\over 2}}\partial^2 F
- {1\over 8}F^{-{5\over 2}}(\partial F)^2 \\
(\nabla\Phi)^2 &=& {1\over 16}F^{-{5\over 2}}(\partial F)^2
\eea
These expressions, and Eq.~\ref{eq:ricscalar} for the Ricci scalar,
indeed satisfy the dilaton equation Eq.~\ref{eq:dilaton}.

\paragraph{Temporal part of the Einstein equation:} 
A short calculation that uses Eq.~\ref{eq:gmnident} and 
Eq.~\ref{eq:iden2} gives:
\be
{1\over 24}F^2_4 \equiv {1\over 24}F^{KLMN} F_{KLMN}=
F^{-2}\partial^2 F - F^{-3} (\partial F)^2
\ee
and therefore:
\be
{1\over 12}e^{2\Phi}(F_0^{~KLM}F_{0KLM}-
{1\over 8}g_{00}F^{KLMN} F_{KLMN})= 
{1\over 48}e^{2\Phi}g_{00}F^2_4
= -{1\over 4}[F^{-2}\partial^2 F - F^{-3} (\partial F)^2 ]
\ee
We must also calculate the covariant derivatives
of the dilaton:
\be
-2\nabla_0 \nabla_0 \Phi = 2\Gamma^k_{00}\partial_k\Phi
= -{1\over 8}F^{-3} (\partial F)^2
\ee
These expressions, and the $R_{00}$ from Eq.~\ref{eq:R00}, satisfy the 
temporal part of the gravition equation
Eq.~\ref{eq:gravity}.

\paragraph{The external part of the Einstein equation:} 
The stress tensor of the 4-form field strength is:
\bea
&~&{1\over 12}e^{2\Phi}(F_i^{~KLM}F_{jKLM}-
{1\over 8}g_{ij}F^{KLMN} F_{KLMN}) \\
&=&-{1\over 4}\delta_{ij} [F^{-1}\partial^2 F -F^{-2} (\partial F)^2]
-{1\over 8}F^{-2}\partial_i F\partial_j F
+{1\over 2}\partial_i g_{\mu\bar{\nu}}\partial_j g^{\mu\bar{\nu}}
\eea
and we also need:
\be
-2\nabla_i \nabla_j\Phi = -2(\partial_i\partial_j-
\Gamma^k_{ji}\partial_k)\Phi
= -{1\over 2}F^{-1}\partial_i\partial_j F +{3\over 4}F^{-2}
\partial_i F\partial_j F -{1\over 8}\delta_{ij}F^{-2}
(\partial F)^2
\ee
The external part of the Einstein equation Eq.~\ref{eq:gravity} can now
be verified by adding these equations, using Eq.~\ref{eq:iden2},
and comparing with the $R_{ij}$ given in Eq.~\ref{eq:Rij}.

\paragraph{The internal part of the Einstein equations:} 
We finally consider the Einstein equation with indices in the
compact direction. The energy momentum carried by the 4-form field
strength is:
\bea
&~&{1\over 12}e^{2\Phi}(F_\mu^{~KLM}F_{\bar{\nu}KLM}-
{1\over 8}g_{\mu\bar{\nu}}F^{KLMN} F_{KLMN}) \\
&=&  [{1\over 8}F^{-{5\over 2}} (\partial F)^2
-{1\over 4}F^{-{3\over 2}}\partial^2 F]g_{\mu\bar{\nu}}
+{1\over 2}F^{-{3\over 2}}\partial F \partial g_{\mu\bar{\nu}}
-{1\over 2}F^{-{1\over 2}}\partial g_{\mu\bar{\rho}}
g^{\bar{\rho}\lambda}\partial g_{\lambda\bar{\nu}}
\eea
and we also need:
\be
-2\nabla_\mu \nabla_{\bar{\nu}}\Phi
= 2\Gamma^k_{\mu\bar{\nu}}\partial_k\Phi
=-{1\over 4}F^{-{1\over 2}}\partial g_{\mu\bar{\nu}}
\ee
Adding these two terms and comparing with $R_{\mu\bar{\nu}}$ in 
Eq.~\ref{eq:Rmn}, we verify the internal part of the 
Einstein equation Eq.~\ref{eq:gravity}.

This completes the verification of the equations of motion.


\begin{thebibliography}{10}

\bibitem{gppkt}
G.~Papadopoulos and P.~K.~Townsend, 
\newblock Intersecting M-Branes.
\newblock {\em Phys. Lett. B}, 379: 273, 1996.
\newblock hep-th/9603087.

\bibitem{sv1}
A.~Strominger and C.~Vafa.
\newblock Microscopic origin of the {B}ekenstein-{H}awking entropy.
\newblock {\em Phys. Lett. B}, 379:99--104, 1996.
\newblock hep-th/9601029.

\bibitem{structure}
F.~Larsen and F.~Wilczek.
\newblock Internal structure of black holes.
\newblock {\em Phys. Lett. B}, 375:37--42, 1996.
\newblock hep-th/9511064.

\bibitem{callan96a}
C.~Callan and J.Maldacena.
\newblock {D}-brane approach to black hole quantum mechanics.
\newblock {\em Nucl. Phys. B}, 472:591--610, 1996.
\newblock hep-th/9602043.

\bibitem{strom96b}
G.T. Horowitz and A.~Strominger.
\newblock Counting states of near-extremal black holes.
\newblock {\em Phys. Rev. Lett}, 77:2369--2371, 1996.
\newblock hep-th/9602051.

\bibitem{dvv1}
R.~Dijkgraaf, E.~Verlinde, and H.~Verlinde.
\newblock {BPS} spectrum of the five-brane and black hole entropy.
\newblock {\em Nucl. Phys. B}, 486:77--88, 1996.
\newblock hep-th/9603126.

\bibitem{bdl}
M.~Berkooz, M.R. Douglas, and R.G. Leigh.
\newblock Branes intersecting at angles.
\newblock {\em Nucl. Phys.B}, 480:265--278, 1996.
\newblock hep-th/9606139.

\bibitem{bl}
V.~Balasubramanian and R.G. Leigh.
\newblock {D}-branes, moduli and supersymmetry.
\newblock {\em Phys. Rev. D}, 55:6415--6422, 1997.
\newblock hep-th/9611165.

\bibitem{gauntlett}
J.~P. Gauntlett, G.~W. Gibbons, G.~Papadopoulos, and P.~K. Townsend.
\newblock Hyper-k\"{a}hler manifolds and multiply intersecting branes.
\newblock hep-th/9702202.

\bibitem{behrndt}
K.~Behrndt and M.~Cvetic.
\newblock {BPS} saturated bound states of tilted p-branes in type {II} string
  theory.
\newblock hep-th/9702205.

\bibitem{mcgill1}
J.C. Breckenridge, G.~Michaud, and R.C. Myers.
\newblock New angles on {D}-branes.
\newblock hep-th/9703041.

\bibitem{costa}
M.~Costa and M.~Cvetic.
\newblock Nonthreshold {D}-brane bound states and black holes with nonzero
  entropy.
\newblock hep-th/9703204.

\bibitem{hambli}
N.~Hambli.
\newblock Comments on {D}irichlet branes at angles.
\newblock hep-th/9703179.

\bibitem{harm}
A.A. Tseytlin.
\newblock Harmonic superposition of {M}-branes.
\newblock {\em Nucl. Phys.B}, 475:149--163, 1996.
\newblock hep-th/9604035.

\bibitem{us}
V.~Balasubramanian, F.~Larsen, and R.G. Leigh.
\newblock Work in progress.

\bibitem{dnotes}
J.~Polchinski.
\newblock {TASI} lectures on {D}-branes.
\newblock hep-th/9611050.

\bibitem{vf}
V.~Balasubramanian and F.~Larsen.
\newblock On {D}-branes and black holes in four dimensions.
\newblock {\em Nucl. Phys.B}, 478:199, 1996.
\newblock hep-th/9604189.

\bibitem{n2area}
K.~Behrndt, G.L. Cardoso, B.~de~Wit, R.~Kallosh, D.~L\"{u}st, and T.~Mohaupt.
\newblock Classical and quantum {N}=2 supersymmetric black holes.
\newblock {\em Nucl. Phys. B}, 488:236--260, 1997.
\newblock hep-th/9610105.

\bibitem{giveon}
A.~Giveon, M.~Porrati, and E.~Rabinovici.
\newblock Target space duality in string theory.
\newblock {\em Phys. Rept.}, 244: 77-202, 1994.
\newblock hep-th/9401139.

\bibitem{ct2}
M.~Cveti\v{c} and A.~Tseytlin.
\newblock Solitonic strings and {BPS} saturated dyonic black holes.
\newblock {\em Phys. Rev. D}, 53:5619--5633, 1996.
\newblock hep-th/9512031.

\bibitem{dkl}
R.R.~Khuri M.J.~Duff and J.X. Lu.
\newblock String solitons.
\newblock {\em Phys. Rep.}, 259:213--326, 1995.
\newblock hep-th/9412184.

\bibitem{cremmer}
E. Cremmer and B. Julia.
\newblock The {SO}(8) supergravity.
\newblock {\em Nucl. Phys.B},159:141, 1979.

\bibitem{hull}
C.~M. Hull and P.~K. Townsend.
\newblock Unity of superstring dualities.
\newblock {\em Nucl. Phys. B}, 438:109, 1995.
\newblock hep-th/9410167.

\bibitem{hull96}
C.~Hull and M.~Cveti\v{c}.
\newblock Black holes and {U}-duality.
\newblock {\em Nucl. Phys. B}, 480:296--316, 1996.
\newblock hep-th/9606193.

\bibitem{kallosh96b}
S.Ferrara and R.~Kallosh.
\newblock Universality of supersymmetry attractors.
\newblock {\em Phys. Rev. D}, 54:1525--1534, 1996.
\newblock hep-th/9603090.

\bibitem{kallosh96a}
R.~Kallosh and B.~Kol.
\newblock E(7) symmetric area of the black hole horizon.
\newblock {\em Phys. Rev. D}, 53:5344--5348, 1996.
\newblock hep-th/9602014.

\bibitem{vfscatter}
V.~Balasubramanian and F.~Larsen.
\newblock Relativistic brane scattering.
\newblock hep-th/9703039, to appear in Nucl. Phys. B.

\bibitem{kt}
I.~Klebanov and A.~Tseytlin.
\newblock Intersecting {M}-branes as four-dimensional black holes.
\newblock {\em Nucl. Phys.B}, 475:179--192, 1996.
\newblock hep-th/9604166.

\bibitem{juan1}
J.~Maldacena.
\newblock {N}=2 extremal black holes and intersecting branes.
\newblock hep-th/9611163.

\bibitem{behrndt2}
K.~Behrndt and T.~Mohaupt.
\newblock Entropy of {N}=2 black holes and their {M}-brane description.
\newblock hep-th/9611140.

\bibitem{gsw}
M.B. Green, J.H. Schwarz, and E.~Witten.
\newblock {\em Superstring Theory}.
\newblock Cambridge University Press, 1987.

\bibitem{mathur96}
S.~Das and S.~Mathur.
\newblock Excitations of {D}-strings, entropy and duality.
\newblock hep-th/9601152.

\bibitem{susskind96}
J.~Maldacena and L.~Susskind.
\newblock {D}-branes and fat black holes.
\newblock hep-th/9604042.

\bibitem{dmvv}
R.~Dijkgraaf, G.~Moore, E.~Verlinde, and H.~Verlinde.
\newblock Elliptic genera of symmetric products and second quantized strings.
\newblock hep-th/9608096.

\bibitem{dvv3}
R.~Dijkgraaf, E.~Verlinde, and H.~Verlinde.
\newblock Counting dyons in {N=4} string theory.
\newblock hep-th/9607026.

\end{thebibliography}

\end{document}